\pdfoutput=1
\documentclass[letterpaper,twocolumn,10pt]{article}
\usepackage{usenix2019,epsfig,endnotes}
\usepackage{graphicx}
\usepackage{subfigure}
\usepackage{tablefootnote}
\usepackage{booktabs}
\usepackage{fancyvrb}
\usepackage{multirow,array,url}
\usepackage[misc,geometry]{ifsym}
\usepackage{bold-extra}
\usepackage{pifont}
\usepackage{amsmath}
\usepackage{tabto}
\usepackage{hyperref}
\usepackage{algorithm}
\usepackage{verbatim}
\usepackage{algpseudocode}
\usepackage[T1]{fontenc}
\usepackage{todonotes}
\usepackage{amsfonts}
\usepackage{soul}
\usepackage{colortbl}
\usepackage{siunitx}
\makeatletter
\g@addto@macro{\UrlBreaks}{\UrlOrds}
\makeatother

\definecolor{cgray}{gray}{0.90}
\definecolor{cgreen}{rgb}{0,0.6,0}
\definecolor{cmauve}{rgb}{0.58,0,0.82}

\begin{document}
%
%
\date{}
\title{\Large \bf Undermining User Privacy on Mobile Devices Using AI}
\hypersetup{pdfauthor={G\"{u}lmezo\u{g}lu et al.},pdftitle={Undermining User Privacy on Mobile Devices Using AI}}
\author{
{\rm Berk Gulmezoglu}\\
Worcester Polytechnic Institute\\
bgulmezoglu@wpi.edu
\and
{\rm\hspace{1.5em}Andreas Zankl}\\
\hspace{1.5em}Fraunhofer AISEC\\
\hspace{1.5em}andreas.zankl@aisec.fraunhofer.de
\and
{\rm\hspace{-1.4em}M. Caner Tol}\\
\hspace{-1.4em}Middle East Technical University\\
\hspace{-1.4em}caner.tol@metu.edu.tr
\and
{\rm\hspace{1.9em}Saad Islam}\\
\hspace{1.9em}Worcester Polytechnic Institute\\
\hspace{1.9em}sislam@wpi.edu
\and
{\rm\hspace{-1.6em}Thomas Eisenbarth}\\
\hspace{-1.6em}University of L\"{u}beck\\
\hspace{-1.6em}thomas.eisenbarth@uni-luebeck.de
\and
{\rm\hspace{1.6em}Berk Sunar}\\
\hspace{1.6em}Worcester Polytechnic Institute\\
\hspace{1.6em}sunar@wpi.edu
}
\maketitle

%
%
\begin{abstract}
  Over the past years, literature has shown that attacks exploiting the microarchitecture of modern processors pose a serious threat to the privacy of mobile phone users.
  This is because applications leave distinct footprints in the processor, which can be used by malware to infer user activities.
  In this work, we show that these inference attacks are considerably more practical when combined with advanced AI techniques.
  In particular, we focus on profiling the activity in the last-level cache (LLC) of ARM processors. 
  We employ a simple Prime+Probe based monitoring technique to obtain cache traces, which we classify with Deep Learning methods including Convolutional Neural Networks.
  We demonstrate our approach on an off-the-shelf Android phone by launching a successful attack from an unprivileged, zero-permission App in well under a minute.
  The App thereby detects running applications with an accuracy of 98\% and reveals opened websites and streaming videos by monitoring the LLC for at most 6 seconds.
  This is possible, since Deep Learning compensates measurement disturbances stemming from the inherently noisy LLC monitoring and unfavorable cache characteristics such as random line replacement policies.
  In summary, our results show that thanks to advanced AI techniques, inference attacks are becoming alarmingly easy to implement and execute in practice.
  This once more calls for countermeasures that confine microarchitectural leakage and protect mobile phone applications, especially those valuing the privacy of their users.
\end{abstract}

%
%
\section{Introduction}
\label{sec:intro}

Today, more than 1 billion people use Android applications~\cite{GoogleAnnouncement}.
The security and privacy of these applications are therefore of great relevance.
The Android OS therefore employs a variety of protection mechanisms.
Apps run in sandboxes, inter-process communication is regulated, and users have some degree of control via the permission system.
The majority of these features protects against software-based attacks and logical side-channel attacks.
The processor hardware, however, also constitutes an attack surface.
In particular, the shared processor cache heavily speeds up the execution of applications.
As a side effect, each application leaves a footprint in the cache that can be profiled by others.
These footprints, in turn, contain sensitive information about the application activity.
Jana et al.~\cite{JanaShmatikov2012} showed that browsing activity yields unique memory footprints that allow to infer accessed web pages.
Oren et al.~\cite{OrenEtAl2015} demonstrated that these footprints can be observed in the cache even from JavaScript code distributed by a malicious website. 
While these attacks have succeeded based on a solid amount of engineering, the increasing complexity of applications, operating systems (OS), and processors make their implementation laborious and cumbersome.
Yet, studying side-channel attacks is important to protect security and privacy critical applications in the long term.
We believe that machine learning techniques, especially Deep Learning (DL), can help make side-channel analysis significantly more scalable.
DL thereby reduces the human effort by efficiently extracting relevant information from noisy and complex side-channel observations.
At the same time, DL introduces a new risk as attacks become more potent and easier to implement in practice.

In this work, we demonstrate this risk and compile a malicious Android application, which, despite having no privileges or permissions, can infer private user activities across application and OS boundaries.
With the App, we are able to detect other running applications with high confidence.
With this information, we focus on activities that happen within an application.
We detect visited websites in Google Chrome and identify videos that are streamed in the Netflix and Youtube applications.
Those inferences are possible by analyzing simple last-level cache (LLC) observations of at most 6 seconds with advanced DL algorithms.
The entire attack succeeds in well under a minute and reveals sensitive information about the mobile phone user.
None of the currently employed protection mechanisms prevent our attack, as the LLC is shared between different processes and can be monitored from user space.
Our cache profiling technique is based on the Prime+Probe (P+P) attack~\cite{TromerEtAl2010}, which relies on cache eviction to monitor certain cache sets.
In contrast to previous work, we implement this eviction with a dynamic eviction set test that succeeds even for imprecise timing sources, random line replacement policies, and missing physical memory addresses information.
This comes at the cost of measurement accuracy and introduces a certain amount of noise in the cache observations.
We counter this effect by applying machine learning to the observations, and compare classical algorithms to advanced deep learning along the way.
While Support Vector Machines (SVMs) and Stacked AutoEncoders (SAEs) struggle during the classification, Convolution Neural Networks (CNNs), a DL technique, succeed in efficiently extracting distinct features and classifying the observations.
As CNNs have recently gained attention in the field of side-channel analysis, we explain our parameter selection and compare it to related work.
For the implementation of our attack, we neither require the target phone to be rooted nor the malicious application to have certain privileges or permissions.
On our test device, a Nexus 5X, the Android OS is up-to-date and all security patches are installed.
The malicious code runs in the background, requires no human contribution during the attack, and draws little attention due to the short profiling phase.

\bigskip
\noindent
{\bf Our Contribution.} In summary, we

\begin{itemize}
  \item propose an inference attack on mobile devices that works without privileges, permissions, or access to special interfaces.
  \item find eviction sets with a novel dynamic timing test that works even with imprecise timing sources, random line replacement policies, and virtual addresses only.
  \item classify cache observations using ML/DL techniques (SVMs, SAEs, CNNs) and thereby infer running applications, opened websites, and streaming videos.
  \item achieve classification rates up to 98\% with a profiling phase of at most 6 seconds. The entire attack succeeds in well under one minute.
\end{itemize}
\bigskip

The rest of the paper is organized as follows: Section~\ref{sec:background} provides background information on classical and modern machine learning techniques.
Section~\ref{sec:attack} explains how we profile the cache and conduct the inference attack.
Section~\ref{sec:results} presents the results of our experiments.
Section~\ref{sec:related_work} gives an overview of previous works and compares our results with other techniques.
Section~\ref{sec:conclusion} concludes our work.
\section{Background}
\label{sec:background}

This section briefly explains the basic concepts of the ML/DL techniques that are used in this work.

\subsection{{Support Vector Machines (SVMs)}}

SVMs construct a classifier by mapping training data into a higher dimensional space, where distinct features can efficiently be separated.
This separation is achieved with a linear or non-linear hyperplane that is built based on an error term.
The error term includes a penalty/regularization parameter that often ranges between 0 and 1.
While bounding the parameter in this range makes the optimization process more expensive, it also enables higher success rates.
For our practical experiments, we enforce this bound and rely on the SVM implementations in LibSVM~\cite{chang2011libsvm}.

\subsection{{Stacked AutoEncoders (SAEs)}}

An AutoEncoder (AE) is a type of neural network that can be trained to reconstruct an input.
The network consists of two parts: an encoder function $h=f(x)$ that extracts distinct features from the input $x$ and a decoder unit that reconstructs the original data $r=g(h)$.
The network is trained such that the error between $r$ and $x$ is minimized.
Stacked AEs are then constructed by combining multiple AEs sequentially.
Their goal is to learn only useful input features instead of learning the exact copy of the input.
This technique is similar to Principal Component Analysis (PCA), which finds a low-dimensional representation of input data in an unsupervised manner.
The hidden layers of SAEs, which learn the important features, have non-linear activation functions and their weights are updated through the backpropagation algorithm.
With a Softmax layer as the final stage, SAEs derive output labels and can thus be used for supervised learning.

\subsection{{Convolutional Neural Networks (CNNs)}}

\begin{figure}[t!]
  \centering
  \includegraphics[width=0.46\textwidth]{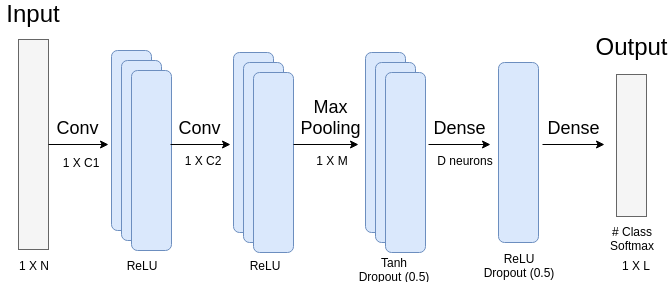}
  \caption{Typical structure of a CNN.}
  \label{fig:CNN}
  \vspace*{2em}
\end{figure}

CNNs are a so-called Deep Learning technique.
They have received broad interest in the field of supervised learning, especially for complex tasks such as image and language classification~\cite{ciresan2011flexible,kim2014convolutional}.
A typical CNN is made up of neurons, which are interconnected and grouped into layers.
Each neuron computes a weighted sum of its inputs using a (non-) linear activation function.
Those inputs stem from the actual inputs to the network or from previous layers.
A typical CNN comprises multiple layers of neurons, as shown in Figure~\ref{fig:CNN}:

\paragraph{Convolutional Layer.}
This layer is the core block of the CNN.
It consists of learnable filters that are slid over the width and height of the input to learn any two-dimensional patterns.
The activation functions in the neurons thereby extract important features from each input.
For efficiency, the neurons are connected only to a local region of the input.

\noindent There are three hyper-parameters that control the behavior of a Convolutional layer.
The first one is called \textbf{depth}.
It defines the number of filters in the layer, which learn different patterns in the input after every iteration.
If all neurons in a single depth have the same weights, the layer can be computed as a convolution of the neurons' weights.
The second parameter is the \textbf{stride}.
It defines how fast the filters are moved over the input.
The most common stride value is 1, which means the filters are moved by one feature at a time.
The last parameter is the \textbf{zero-padding}.
The benefit of zero-padding is to control the spatial size of the output, allowing to match the input and output size of the layer.

\paragraph{Pooling Layer.}
After the Convolutional layer, networks typically add a Pooling layer to reduce the spatial size of the data representation.
It helps to decrease the number of parameters and computation in the network, and also reduces the risk of over-fitting.
The Pooling layer resizes the input spatially, using the $max$ operation.
Basically, the Pooling layer applies a filter (typically 2$\times$2) to the input and the maximum coefficient from every subregion is stored. 
Therefore, only the most dominant features from the Convolution layer are conveyed to the next layer.  

\paragraph{Fully Connected/Dense Layer.}
After detecting high level features in previous layers, a Dense layer is added to the end of the network.
The neurons in this layer are connected to all the activations in the previous layer.
Thus, additional relations between activations can be learned.
If this layer is the final one before the output, then a Loss layer is implemented.
In the Loss layer, the output size is set to $1 \times L$, where $L$ is the number of predicated labels or classes.
Then, the loss function is computed by using \textit{Softmax} or \textit{Sigmoid cross-entropy} to evaluate the performance of the trained model.
Finally, the loss value is decreased by implementing a backpropagation algorithm that updates the weights in the previous layers.
After the test data is classified, a \textit{probability estimate} (confidence ratio) is provided by the model.
The probability estimate indicates the confidence of the model for a predicted class. 
For each class, a probability estimate is given and the sum of all estimates equals to 1.

\section{Inference Attack}
\label{sec:attack}

This section discusses the threat model and the steps of our proposed inference attack.
Subsequently, details about our cache profiling technique are given, including finding eviction sets and post-processing measurements.

\subsection{Threat Model}

In our threat model, we assume that a mobile device user installs a malicious application from the Google Play store on Android.
Malicious Apps frequently offer benign functionality to disguise malicious background activities (e.g. hidden crypto currency mining).
The malicious code needed for our attack operates from user space and does not need any App permissions.
This means that we neither require a rooted phone, nor ask the user for certain permissions, nor rely on any exploits, e.g. to escalate privileges or to break out of sandboxes.
Furthermore, we do not rely on features that might not be available on all Android phones.
This includes the use of memory pages that are larger than \,4\,KB.
The sole task of our malicious code is to profile the LLC and classify victim activities with pre-trained ML/DL models.
Once, the LLC profiles have been gathered, the models are queried to infer sensitive information.

\subsection{Attack Outline}

\begin{figure}[t!]
  \centering
  \includegraphics[width=0.48\textwidth]{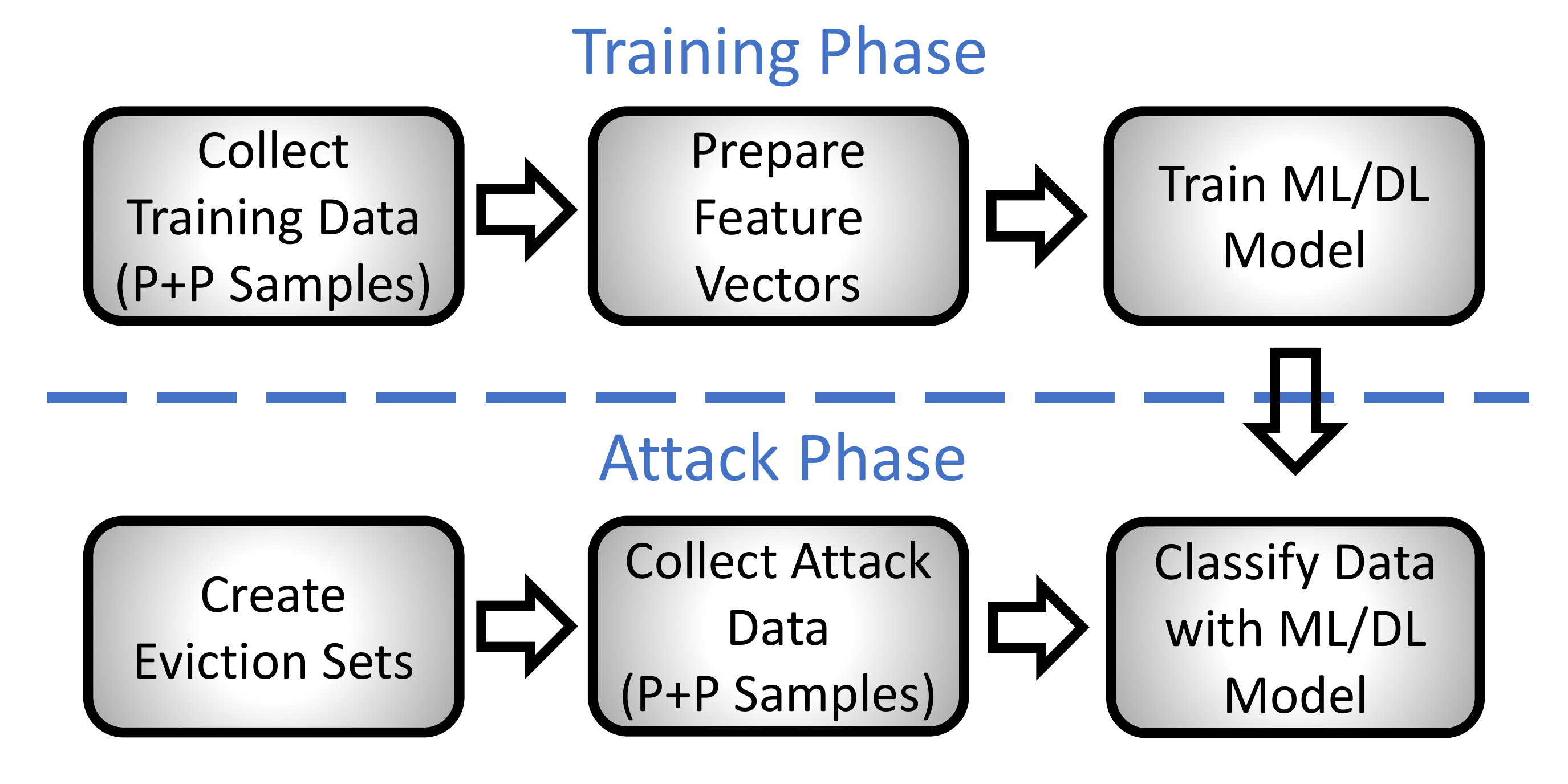}
  \caption{Steps of the proposed inference attack.}
  \label{fig:process}
\end{figure}

The proposed inference attack is conducted in two main phases, the steps of which are visualized in Figure~\ref{fig:process}.
In the \textbf{training phase}, the attacker creates ML/DL models on a training device that is similar to the target device.
Most importantly, the processor should be identical on both devices.
The ML/DL models are created by recording raw LLC profiles of target applications, websites, and videos, followed by preparing the feature vectors, and training the ML/DL algorithms with them.
The trained models are then directly integrated into the malicious application, which is subsequently published in the App store.
In the \textbf{attack phase}, the malicious App prepares eviction sets for profiling the LLC on the target device.
Subsequently, the LLC sets are profiled in a Prime+Probe manner and the feature vectors are extracted.
Finally, the LLC profile is classified with the pre-trained models to extract opened applications, websites, and streamed videos.
All steps of the attack phase are lightweight and can be executed in the background without drawing notable attention.

\subsection{Finding Eviction Sets}
\label{subsec:evsetcrea}

Once deployed, the first task of the malicious App is to find \textit{eviction sets} on the target device.
An eviction set is essentially a group of memory addresses that map to the same cache set.
Eviction sets are useful to profile the LLC and therefore to capture the cache activity of other applications.
A recent study by Vila et al.~\cite{VilaKoepf2018} investigates the problem of finding eviction sets.
The authors focus their evaluation on Intel and do not further analyze ARM processors.
In this work, however, we target ARM-based mobile devices.
Nevertheless, the authors give a comprehensive overview of previous search algorithms and we therefore refer the interested reader to this study for further information.
Algorithms relevant to this work are discussed in the paragraph below.

Lipp et al.~\cite{LippEtAl2016} derive eviction sets from physical address information which they obtain from the \textit{pagemap} file that is typically found under \textit{/proc/self/pagemap} on Linux systems.
After their work was published, Android restricted the access to \textit{pagemap} entries from user space\footnote{\scriptsize\url{https://source.android.com/security/bulletin/2016-03-01}}.
Liu et al.~\cite{LiuEtAl2015}, Irazoqui et al.~\cite{IrazoquiEtAl2015}, and Gruss et al.~\cite{gruss2016rowhammer} rely on \textit{huge pages}, i.e. memory pages with typical sizes of >\,1\,MB, to obtain those physical address bits that reveal the cache set numbers.
Oren et al.~\cite{OrenEtAl2015} and Bosman et al.~\cite{bosmanetal2016} rely on special page allocation mechanisms in web browsers and operating systems that simplify the eviction set search.
Genkin et al.~\cite{GenkinEtAl2018} build eviction sets from sandboxed code within a web browser, while only relying on virtual address information.
Yet, they still require a precise and low-noise timing source to distinguish cache hit and miss.
In contrast to these previous works, we neither rely on physical address information (whether obtained from pagemap, huge pages, or elsewhere), nor on certain features of memory allocators, nor on a precise timing source.
Our approach for finding eviction sets is purely based on virtual addresses and robust against imprecise and noisy timing sources.
In addition, we found our approach to be resilient against the random line replacement policy implemented in many ARM processors.
The majority of previous work, except for~\cite{LippEtAl2016}, studied Intel and AMD processors, which typically implement different replacement policies.

\begin{algorithm}[!t]
  \begin{algorithmic}[1]
    \caption{Finding Eviction Sets.}
    \label{alg:evict}
    \State $\mathcal{T} = \{\}$
    \State $t_{old} = 0$
    \For {$i$ \textbf{from} $1$ \textbf{to} $n$}
    \State $add\left(\mathcal{T}, i\right)$
    \State $t_{\mathcal{T}} = access\left(\mathcal{T},\,r\right)$
    \If {$\left(t_{\mathcal{T}} - t_{old}\right) > \tau_{jump}$}
    \State $\mathcal{E} \leftarrow \{\}$
    \ForAll {$p$ in $\mathcal{T}$}
    \State $t_{p} = access\left(\mathcal{T}\textbackslash\{p\},\,r\right)$
    \If {$\left(t_{\mathcal{T}} - t_{p}\right) > \tau_{jump}$}
    \State $add\left(\mathcal{E},\,p\right)$
    \EndIf
    \EndFor
    \State $report\left(\mathcal{E}\right)$
    \State $remove\left(\mathcal{T},\,\mathcal{E}\right)$
    \State $t_{old} = access\left(\mathcal{T},\,r\right)$
    \Else
    \State $t_{old} = t_{\mathcal{T}}$
    \EndIf
    \EndFor
  \end{algorithmic}
\end{algorithm} 

\begin{figure}[t!]
  \centering
  \subfigure[Average page access times, $t_{\mathcal{T}}$, for an increasing number of pages in $\mathcal{T}$.]{\includegraphics[width=0.48\textwidth]{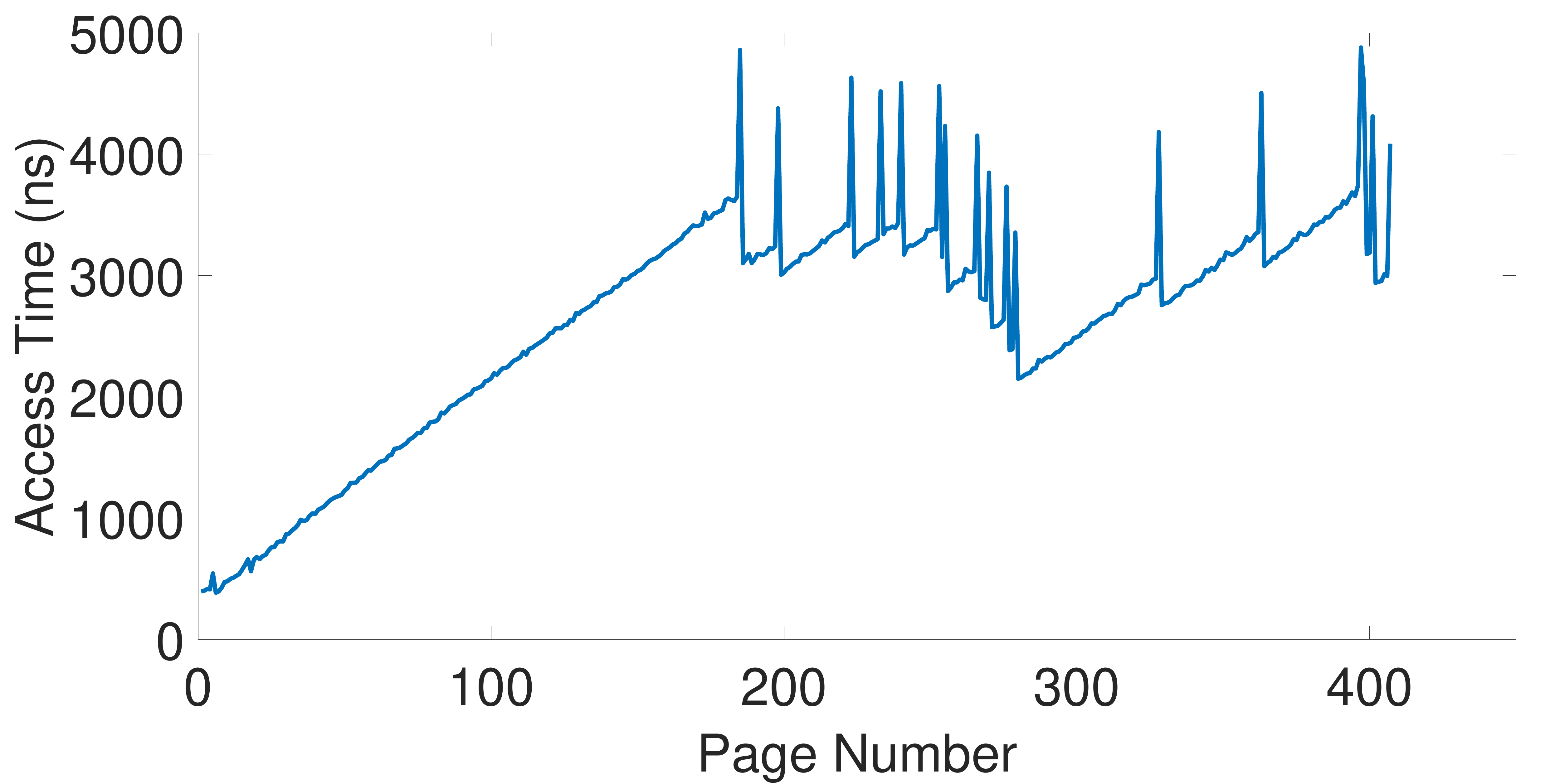}\label{fig:evict}}
  \subfigure[Average page access times, $t_{p}$, used to filter an eviction set from $\mathcal{T}$.]{\includegraphics[width=0.48\textwidth]{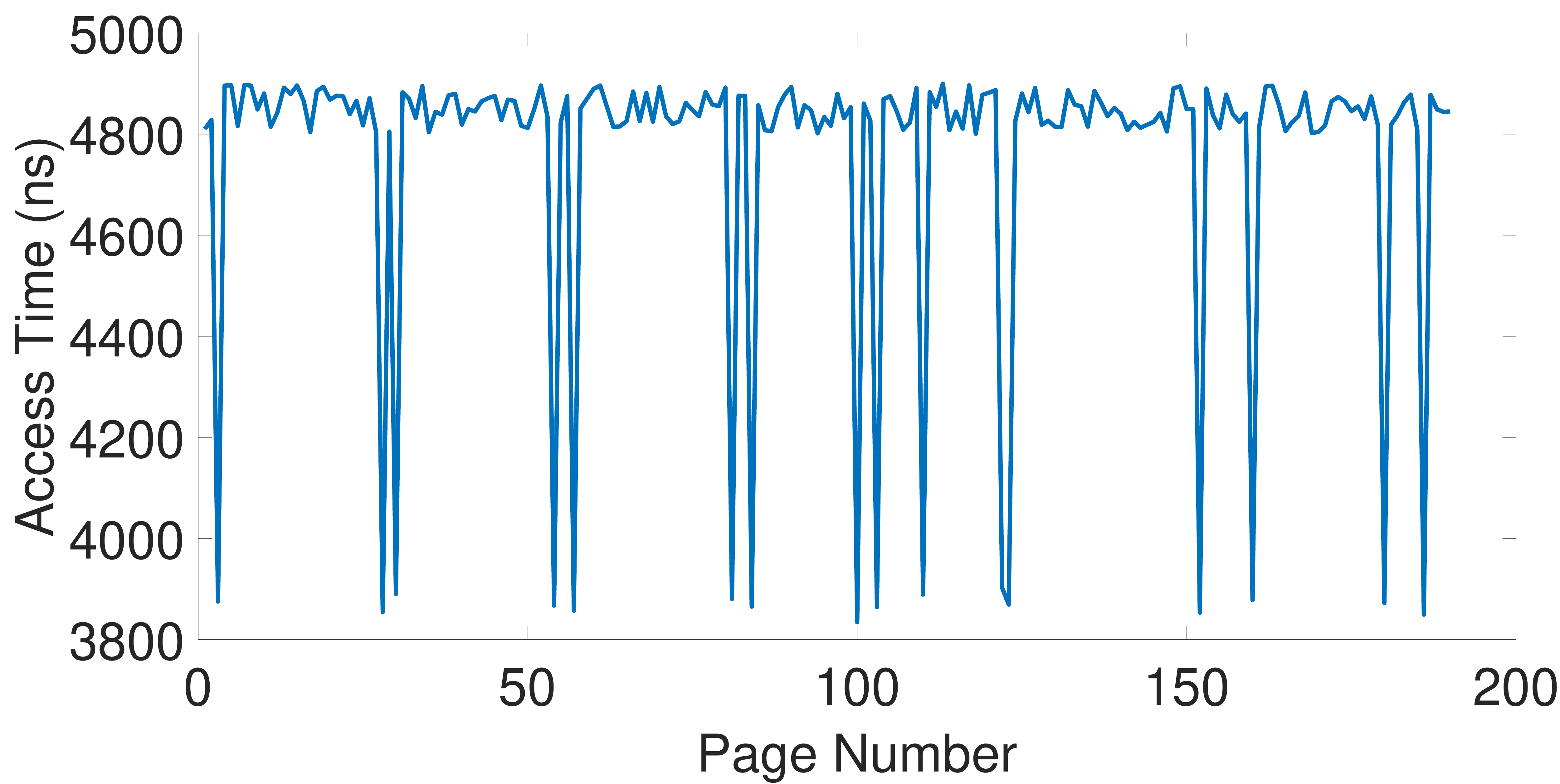}\label{fig:candidate}}
  \caption{Plots of (a) $t_{\mathcal{T}}$ and (b) $t_{p}$, as used in Algorithm~\ref{alg:evict}.}
  \label{fig:eviction}
\end{figure}

Algorithm~\ref{alg:evict} outlines our dynamic timing test that attributes memory addresses to cache sets.
Prior to the test, we assume that $n$ memory pages have been requested and are available as a memory pool.
Note that we don't pose any requirements on the memory page size, and as such, our algorithm works with pages of \,4\,KB and smaller.
Since we want to evict the entire last-level cache, we need to choose $n$ such that the requested memory area is large enough to fill it.
In our experiments, we request a memory area that is twice as large as the LLC.
This turned out to be sufficient for deriving all eviction sets.
Algorithm~\ref{alg:evict} iterates over the allocated memory pages in sequential order.
Each unused page is first added to a temporary eviction set $\mathcal{T}$ (line 4).
Next, the first byte of each page in $\mathcal{T}$ is accessed and the average time $t_{\mathcal{T}}$ of this access cycle is measured.
The parameter $r$ determines how often the access cycle is repeated.
In each cycle, all pages in $\mathcal{T}$ are accessed once.
The overall timing is then divided by $r$ to obtain the average.
This is useful to account for different replacement policies and imprecise timing sources.
A detailed discussion of $r$ is given later in this section.
The access time $t_{\mathcal{T}}$ is then compared with the time from the previous loop cycle (line 6), where $\mathcal{T}$ was one page smaller.
If the time difference is higher than a threshold $\tau_{jump}$, then there is a systematic contention in a cache set.
In other words, the pages in $\mathcal{T}$ entirely fill one set and cause a line replacement in the process.
This is illustrated in Figure~\ref{fig:evict}, which shows the average access time $t_{\mathcal{T}}$ over an increasing number of pages in $\mathcal{T}$.
As long as no set contention occurs, the average timings increase steadily.
Once a contention happens, the timing peaks.
Each peak in the plot indicates that one cache set is completely filled.

After a set contention is detected, Algorithm~\ref{alg:evict} iterates over all pages in $\mathcal{T}$, temporarily excludes them from $\mathcal{T}$, accesses this reduced set, and stores the average access time in $t_{p}$ (line 9).
If the time difference between $t_{\mathcal{T}}$ and $t_{p}$ is again bigger than $\tau_{jump}$, then $p$ belongs to the eviction set.
This is illustrated in Figure~\ref{fig:candidate}, which shows the average access time $t_{p}$ for all candidate pages $p$ in $\mathcal{T}$.
As soon as a candidate is part of the eviction set, the systematic set contention vanishes and the access time $t_{p}$ drops.
Each drop in the plot therefore indicates a member of the eviction set.
Those pages are then added to the final eviction set $\mathcal{E}$, which is reported on line 14.
The entries in $\mathcal{E}$ are subsequently removed from $\mathcal{T}$, before the outer loop continues to add unused pages to $\mathcal{T}$.
Once the outer loop reaches $n$, we derive the eviction sets for all LLC sets from the $\mathcal{E}$ that were reported by Algorithm~\ref{alg:evict}.
The number of reported sets $\mathcal{E}$ is defined as $m$.

\paragraph{\textbf{Eviction Set Derivation and Duplicates.}}
Each reported eviction set $\mathcal{E}$ contains a list of memory pages, of which the first byte maps to the same cache set.
But this is not all we get from a given set $\mathcal{E}$.
Since one memory page typically fits more than one cache line, we can derive multiple evictions sets from one $\mathcal{E}$.
With 4\,KB pages and 64-byte cache lines, for example, there are 64 lines on one page.
Since memory pages are chunks of contiguous physical memory, we know that those 64 lines belong to 64 consecutive cache sets.
Hence, we can derive 64 adjacent eviction sets from one $\mathcal{E}$ (the first being $\mathcal{E}$ itself) by simply adding multiples of 64 to the start address of the pages.
Depending on how large we chose $n$, it can happen that Algorithm~\ref{alg:evict} reports multiple eviction sets that map to the same cache sets.
We therefore need to check all reported sets for duplicates and remove them.
This procedure is outlined in Algorithm~\ref{alg:duplicate}.
It starts by storing the indices of all $m$ reported eviction sets from Algorithm~\ref{alg:evict} in the list $\mathcal{B}$ for bookkeeping (line 2).
As long as $\mathcal{B}$ is not empty, the first index is removed and assigned to $f$ (line 4).
The algorithm then iterates over all remaining indices $s$ and accesses the corresponding eviction sets $\mathcal{E}_{f}$ and $\mathcal{E}_{s}$.
In particular, one page in $\mathcal{E}_{f}$ and all pages in $\mathcal{E}_{s}$ are accessed consecutively, and the whole process is repeated $r$ times.
If the average timing $t_{\mathcal{E}}$ of these access cycles is larger than a threshold $\tau_{jump}$, then $\mathcal{E}_{f}$ and $\mathcal{E}_{s}$ map to the same cache set.
If this happens, the affected index $s$ is removed from $\mathcal{B}$ (line 8), and the iteration continues.
After all eviction sets have been tested, the index in $f$ is added to the final list $\mathcal{F}$ (line 11).
With each loop, $\mathcal{B}$ is shrinking as duplicate eviction sets are removed.
Once $\mathcal{B}$ is empty, $\mathcal{F}$ contains the list of unique eviction set indices.

\begin{algorithm}[!t]
  \begin{algorithmic}[1]
    \caption{Removing Duplicates.}
    \label{alg:duplicate}
    \State $\mathcal{F} = \{\}$
    \State $\mathcal{B} = \{1..m\}$
    \While {$notempty\left(\mathcal{B}\right)$}
    \State $f = pop\left(\mathcal{B}\right)$
    \For {$s$ in $\mathcal{B}$}
    \State $t_{\mathcal{E}} = access\left(onepage\left(\mathcal{E}_{f}\right)\,+\,\mathcal{E}_{s},\,r\right)$
    \If {$\left(t_{\mathcal{E}} > \tau_{jump}\right)$}
    \State $remove\left(\mathcal{B},\,s\right)$
    \EndIf
    \EndFor
    \State $add\left(\mathcal{F},\,f\right)$
    \EndWhile
  \end{algorithmic}
\end{algorithm} 

\paragraph{\textbf{Timer Precision and Noise.}}
Both algorithms \ref{alg:evict} and \ref{alg:duplicate} are designed to compensate imprecise and noisy timing sources.
Although previous work~\cite{GenkinEtAl2018,SchwarzEtAl2017} suggests that accurate timers can be crafted even in environments that restrict access to high-precision timing sources, this engineering effort can be saved for our inference attack.
We believe this adds to the practicality of our attack and shows that inference attacks are a realistic threat.
The precision and noise compensation in our algorithms is done by tuning the parameter $r$, as well as the threshold $\tau_{jump}$.
$r$ defines the number of access cycles, i.e., how often a selection of memory pages is accessed.
$\tau_{jump}$ defines how the timing of these access cycles should be evaluated.
In Algorithm~\ref{alg:evict}, the accesses on line 5 will typically cause cache hits until the gathered pages trigger a systematic set contention.
The difference between $t_{\mathcal{T}}$ and $t_{old}$ will therefore be in the order of $t_{miss}$, where $t_{miss}$ is the duration of a cache miss.
Similarly, the accesses on line 9 will cause cache hits, if the candidate page $p$ is part of the eviction set.
In this case, the difference between $t_{\mathcal{T}}$ and $t_{p}$ will again be around $t_{miss}$.
Therefore, $\tau_{jump}$ can initially be set slightly smaller than $t_{miss}$ and adjusted based on experimental data.
The choice of $r$ depends on the precision of the timer and the measurement quality.
In general, $r$ should be set such that $r\cdot t_{miss}$ is larger than the precision of the timer.
It can be increased further, if high levels of noise are encountered, e.g., due to random line replacement.
The choices for $\tau_{jump}$ and $r$ also hold for Algorithm~\ref{alg:duplicate}, where the accesses on line 6 will typically cause cache hits until $\mathcal{E}_{f}$ and $\mathcal{E}_{s}$ are duplicates.
When this happens, a systematic set contention will occur, as the chosen page from $\mathcal{E}_{f}$ will be evicted by $\mathcal{E}_{s}$.
In our experiments, we set $r$ between 900 and 1000, and $\tau_{jump}$ to 500.
The timer available on our test device provides a precision of 52\,ns.
Although this resolution would not require large values for $r$, we also need to compensate the random replacement policy that is implemented on our target device.

\paragraph{\textbf{Line Replacement Policies.}}
The parameter $r$ causes repetitive accesses to cache lines.
This signals the cache controller that the accessed lines are of heightened interest and should not be replaced.
For least-recently-used (LRU) replacement policies, this is obviously beneficial, as unrelated cache activity will less likely interfere with the eviction set finding.
But also random replacement policies benefit, because averaging over $r$ access cycles attenuates the effect of unintended line replacements that happen due to random line selection.
Our experiments outlined in Section~\ref{sec:results} show that the selection of $r$ as stated above is sufficient to compensate the effects of the random line replacement used on our test device.

\paragraph{\textbf{Implementation and Limitations.}}
Only few requirements have to be satisfied to find eviction sets with our approach.
Memory must be allocated and accessed, and the accesses must be timed.
Memory pages can be of arbitrary size and the timing source can be coarse-grained.
This allows our algorithm to be implemented in user space, and thus in a plethora of environments, including desktop computers, cloud servers, and mobile devices.
Even sandboxes and virtual machines are typically no obstacle, enabling remote attacks, e.g., from JavaScript.
The limitation of our approach is that we don't know the exact mapping of eviction sets to cache sets.
We only know that we have one eviction set for each existing cache set.
However, this is not a deficiency of our approach, but a direct consequence of not using physical address information.
This choice makes our approach more practical and, thanks to the application of advanced ML/DL techniques, still allows successful inference attacks.

\paragraph{\textbf{Performance.}}
We evaluated algorithms \ref{alg:evict} and \ref{alg:duplicate} on our test device, a Nexus 5X, with the parameters stated above.
The targeted LLC is 2\,MB in size and 16-way set-associative.
It implements a random replacement policy and features 64-byte cache lines.
Requested memory pages have a standard size of 4\,KB.
Based on 1000 evaluation runs, algorithms \ref{alg:evict} and \ref{alg:duplicate} successfully yield all eviction sets with an average runtime of 20 seconds.
Note that during our inference attack, eviction sets must be found only once and remain unchanged until the malicious App is restarted.

\subsection{Profiling the Last-level Cache}

We profile the cache activity in the LLC using the unique eviction sets that we obtain from algorithms \ref{alg:evict} and \ref{alg:duplicate}.
For each $\mathcal{E}$, we derive all adjacent eviction sets and use them to Prime+Probe~\cite{TromerEtAl2010} the corresponding cache sets.
We thereby group observations of multiple cache sets, which we call \textit{virtual cache sets}.
Each virtual cache set consists of multiple individual cache sets that all belong to one $\mathcal{E}$ and its derived eviction sets.
Accordingly, these individual cache sets are called \textit{physical cache sets}.
As stated earlier, the exact order in which eviction sets map to physical cache sets is unknown.
Yet, one virtual cache set comprises a number of consecutive physical cache sets.
To profile a virtual cache set, we fill (prime) it with the eviction sets, before re-filling (probe) it a second time immediately afterwards.
The time it takes to probe reflects the activity of other applications in the virtual cache set.
In general, high levels of activity will increase the probing time, whereas low levels will decrease it.
Each probing time constitutes a \textit{measurement sample} of the virtual cache set.
The LLC profile then consists of the measurement samples for all available virtual cache sets.

\subsection{Post-processing and Feature Vectors}
\label{subsec:preproc}

The measured LLC profiles are post-processed before being classified by the ML/DL algorithms.
For the explanation, we assume that each LLC profile contains $n_{T}$ measurement samples per virtual cache set.
The following list outlines the steps, the last of which is applied to only one feature vector.

\begin{itemize}
	\item Eliminate timing outliers with a threshold ($\tau_{O}$).
	All outliers are omitted in the subsequent analysis by replacing them with the sample median.
	In our experiments, $\tau_{O}$ is set to 5\,$\mu$s .	
	
	\item Convert timing samples to binary representation.
	A threshold ($\tau_{H}$) decides whether a sample value is \textit{low} or \textit{high}.
	In our experiments, $\tau_{H}$ is set to 750\,ns.
	
	\item Compress samples by grouping high values.
	All bursts of consecutive high timing samples are reduced to one high value.
	
	\item Transform measurement samples into the frequency domain using a fast Fourier transformation (FFT).
	This has shown to be a simple, yet effective way to compensate noise in the measurements~\cite{gulmezoglu2017cache,OrenEtAl2015}.
\end{itemize}

\noindent After the post-processing steps, we extract meaningful feature vectors from the measurement samples, before feeding everything to the ML/DL algorithms.
In total, we derive three different feature vectors that are outlined in the following paragraphs.
While the first two are based on profiling virtual cache sets, the last one is based on profiling physical (i.e. individual) cache sets.
This feature vector is used to evaluate the performance of our inference attack in Section~\ref{sec:results}.

\paragraph{\textbf{Virtual Cache Sets.}}	
In this case, the feature vector contains the number of high samples for each virtual cache set.
The size of the feature vector is therefore at most $n_{T}$ $\times$ the number of virtual cache sets.

\paragraph{\textbf{Virtual Cache Sets + FFT.}}
In this case, the last post-processing step, the Fourier transformation, is applied to the measurement samples in addition.
The sampling rate $n_{S}$ is derived by dividing 1 second by the duration of one Prime+Probe cycle.
For $n_{T}$ samples, the FFT yields $\frac{n_{T}}{2}$ components, excluding the DC component.
This is because the FFT is symmetrical.
For each virtual cache set, we compute the FFT of all corresponding physical cache sets individually, and then group these components into $n_{fft}$ bins by summing up their absolute bin values.
Thus, each virtual cache set yields $n_{fft}$ frequency components.
The overall feature vector size is therefore $n_{fft}$ $\times$ the number of virtual cache sets.

\paragraph{\textbf{Physical Cache Sets.}}
In this case, the feature vector contains the number of high samples for each individual LLC set.
The cache sets are thereby profiled individually and in order, according to their cache set index.
This requires the knowledge of physical address information, and thus a stronger attacker model than the one we assume for our inference attack.
Consequently, the feature vector is only used in a comparison attack, with which we evaluate the success of our proposed inference attack.
The size of the feature vector is at most $n_{T}$ $\times$ the number of physical cache sets.

\subsection{Attack Evaluation}

As stated above, we evaluate our proposed inference attack by comparing it to a more powerful attack.
In particular, we implement the inference attack with elevated privileges (e.g. root), which allow to query precise timers and obtain exact cache set indices of profiled sets.
The eviction sets are derived using physical address information from \textit{pagemap} entries, instead of using algorithms \ref{alg:evict} and \ref{alg:duplicate}.
This allows to attribute eviction sets to physical cache sets and enables fine-grained LLC profiling.
Consequently, the comparison attack profiles individual cache sets in ascending order and thereby captures application activity with high resolution.
In the rest of the paper, this attack is called \textit{comparison attack}.

\section{Experiment Setup and Results}
\label{sec:results}

This section provides details about our experiment setup, including our target device and ML/DL parameter selection.
Subsequently, the results of the attack evaluation are presented and discussed.
In total, we conduct three experiments, in which our malicious App detects applications (1), identifies visited websites (2), and infers streamed videos (3).
The classification rates in all experiments are based on the most likely label.

\subsection{Target Device and ML Testing}

The target device in our experiments is a Google Nexus 5X with Android v8.0.0.
It features four ARM Cortex-A53 and two ARM Cortex-A57 processor cores.
The malicious code runs on one of the A57 cores and profiles the last-level cache in the background.
The target applications are launched and transition automatically to the A57 processor cluster.
This is because the scheduler assigns resource-hungry processes (e.g. browser or multimedia applications) to the A57 cores to leverage their high performance. 
During all experiments, the system was connected to the campus wireless network and background processes from the Android OS and other Apps were running.
The timing source in our malicious App is the POSIX \textit{clock\_gettime} system call, which is available on most Linux systems as well as on Android.
The inferred applications, websites, and videos are listed in Appendix~\ref{sec:appendix}.
For website inference, we run Google Chrome and for video inference, we run Netflix and Youtube\footnote{Chrome v64.0.3282.137, Netflix v6.16.0, Youtube v13.36.50}.

SVM and SAE classification is implemented with the help of LibSVM~\cite{chang2011libsvm}, whereas CNN classification is done using custom Keras~\cite{chollet2015keras} scripts together with the Tensorflow~\cite{abadi2016tensorflow} GPU backend.
The CNN is trained on a workstation with two Nvidia 1080Ti (Pascal) GPUs, a 20-core Intel i7-7900X CPU, and 64 GB of RAM.
For all classifiers, 90\% of the measured LLC profiles are selected randomly for the training phase, while the rest of the data is chosen to evaluate the efficiency of the trained models.
This 10\% holdout approach yields the classification accuracies that are presented in the subsequent sections.
Throughout the experiments, we observed that 10-fold cross-validation results are consistent with a 10\% holdout approach.
In total, we collected more than 800 GB of cache profiling data to evaluate our inference attack.

\subsection{ML/DL Parameter Selection}

The following paragraphs provide the parameter selection of the ML/DL techniques and further details about their usage.

\paragraph{\textbf{SVM.}}
\textit{Physical} and \textit{virtual cache set} feature vectors are classified with a linear SVM, while the \textit{virtual cache set + FFT} feature vector is classified with a non-linear SVM.
This is because the FFT is computed with non-linear functions (cos, sin) and the labels are linearly increasing for the classes. 
Therefore, a non-linear SVM is chosen to classify the FFT components.
The superior performance of a non-linear SVM is verified in preliminary experiments.
Similarly, we determine that the linear kernel type outperforms radial basis and polynomial options for \textit{virtual cache set} and \textit{virtual cache set + FFT} feature vectors.

\paragraph{\textbf{SAE.}}
The two hidden layers of the SAE comprise 250 and 50 neurons, respectively.
The maximum number of epochs is set to 400, since no improvements can be observed afterwards.
We decrease the effect of over-fitting  by setting the L2 weight regularization parameter to 0.01.
The output layer is a Softmax layer.

\paragraph{\textbf{CNN.}}
The construction of the CNN follows the structure illustrated in Figure~\ref{fig:CNN}, Section~\ref{sec:background}.
The CNN consists of two 1-D Convolution layers that are followed by MaxPooling, Dropout, as well as Flatten and Dense layers.
The selection of the layer parameters is done with the help of preliminary experiments.
Table~\ref{table:parameter_selection} in Appendix~\ref{sec:appendix} shows the parameter space that we explored.
Eventually, we selected the parameters that yielded the lowest validation loss (highlighted in bold).
The size of the first 1-D Convolution layer is varied from 8 to 1024.
The highest classification rate is obtained with a size of 512.
Similarly, the size of the second Convolution layer is varied between 32 and 256, and eventually fixed to 256.
A third Convolution layer does not improve the classification.
The activation function in the Convolution layers is set to Rectified Linear Unit (ReLU).
The size of the subsequent MaxPooling layer is varied from 2 to 8.
The default size of 2 yields the best results.
The dropout of the following Dropout layer is varied between 0.1 and 0.5, and finally set to 0.2.
A higher dropout, as for example used in computer vision, adversely affects the classification.
Next, the kernel size is adjusted, and out of the values between 3 and 27, a size of 9 achieves the highest success rates.
A Flatten layer shapes the data in our network, before a Dense layer of size 200 is appended.
The activation function of the Dense layer is set to hyperbolic tangent (tanh).
Finally, we employ a set of standard choices: the kernel initializers are chosen uniformly at random, an Adam optimizer is used to speed up the training phase, and the batch size is set to 50, as its effect on the classification rate is negligible.

\subsection{Evaluation Results}

The following sub-sections present the evaluation results for application, website, and video inference.

\subsubsection{Application Inference}

For the application inference attack, we profile 100 random mobile applications from the Google Play Store, including dating, political, and spy Apps.
The full list is given in Table~\ref{tbl:appsdet} in Appendix~\ref{sec:appendix}.
The first 70 applications are used to train the ML/DL models, while the remaining 30 Apps are treated as unknown Apps.
Each App is started and profiled for 1.5 seconds as described in Section~\ref{sec:attack}.
Within this time frame, we collect $n_{T} = 96,000$ measurement samples per virtual cache set.
For the FFT computation, the sampling rate $n_{S} = 1.9\,MHz$ and the number of bins $n_{fft} = 15$.

\begin{figure}[!t]
  \centering
  \subfigure[Physical Cache Sets.]{\includegraphics[width=0.48\textwidth,clip,trim=0 2cm 0 0]{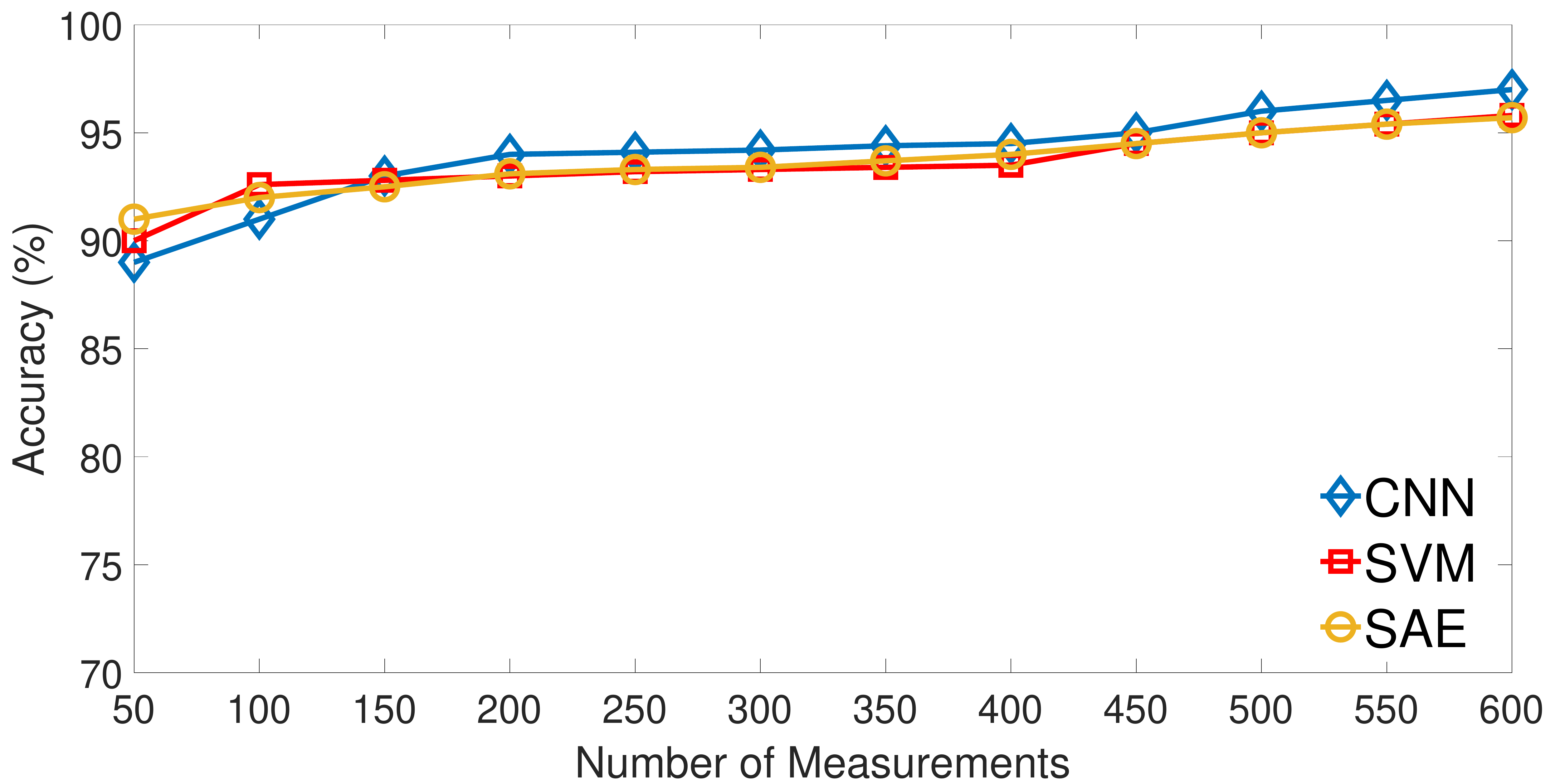}\label{fig:ACM}}
  \subfigure[Virtual Cache Sets.]{\includegraphics[width=0.48\textwidth,clip,trim=0 2cm 0 0]{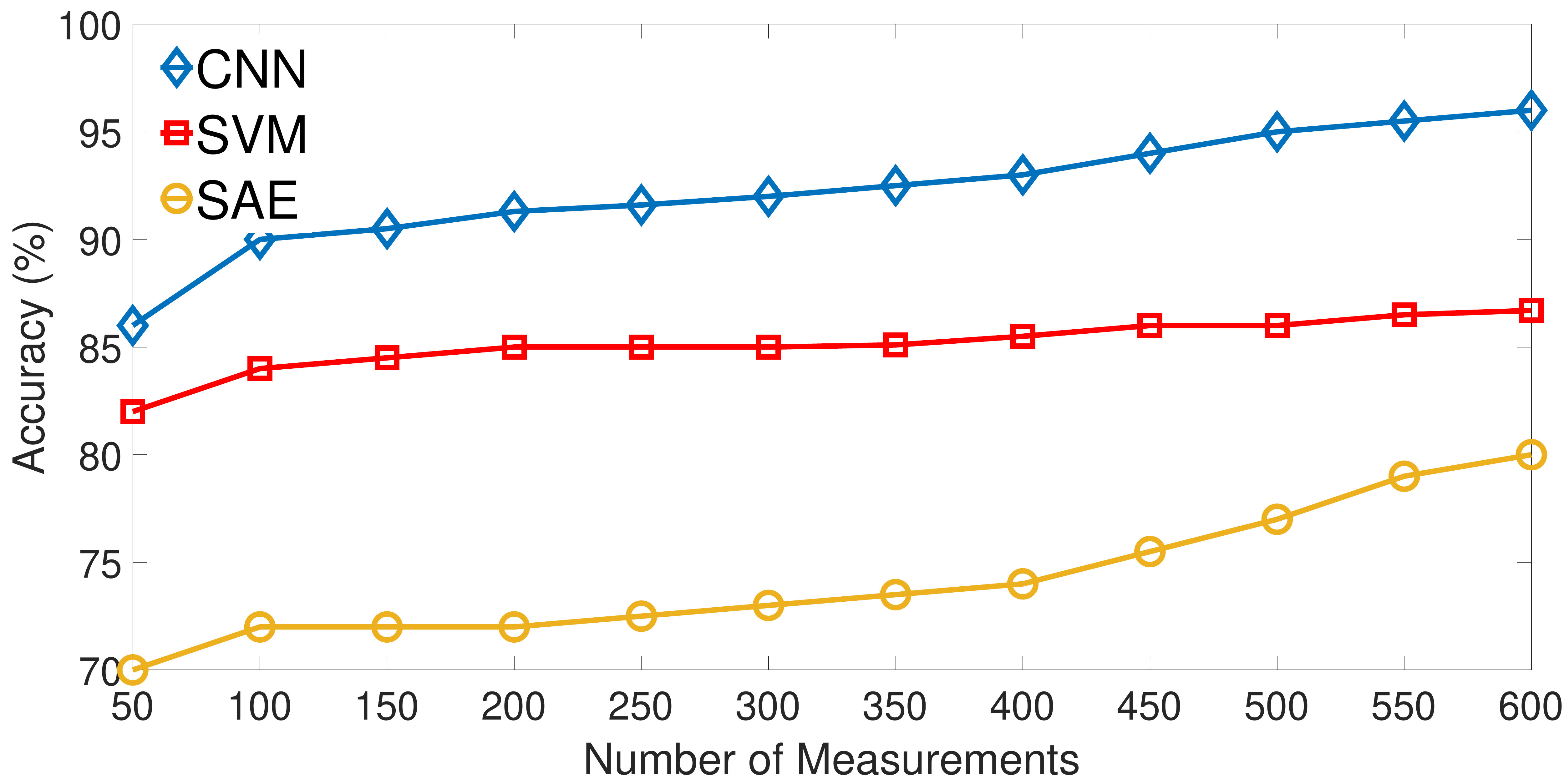}\label{fig:ACMT}}
  \subfigure[Virtual Cache Sets + FFT.]{\includegraphics[width=0.48\textwidth]{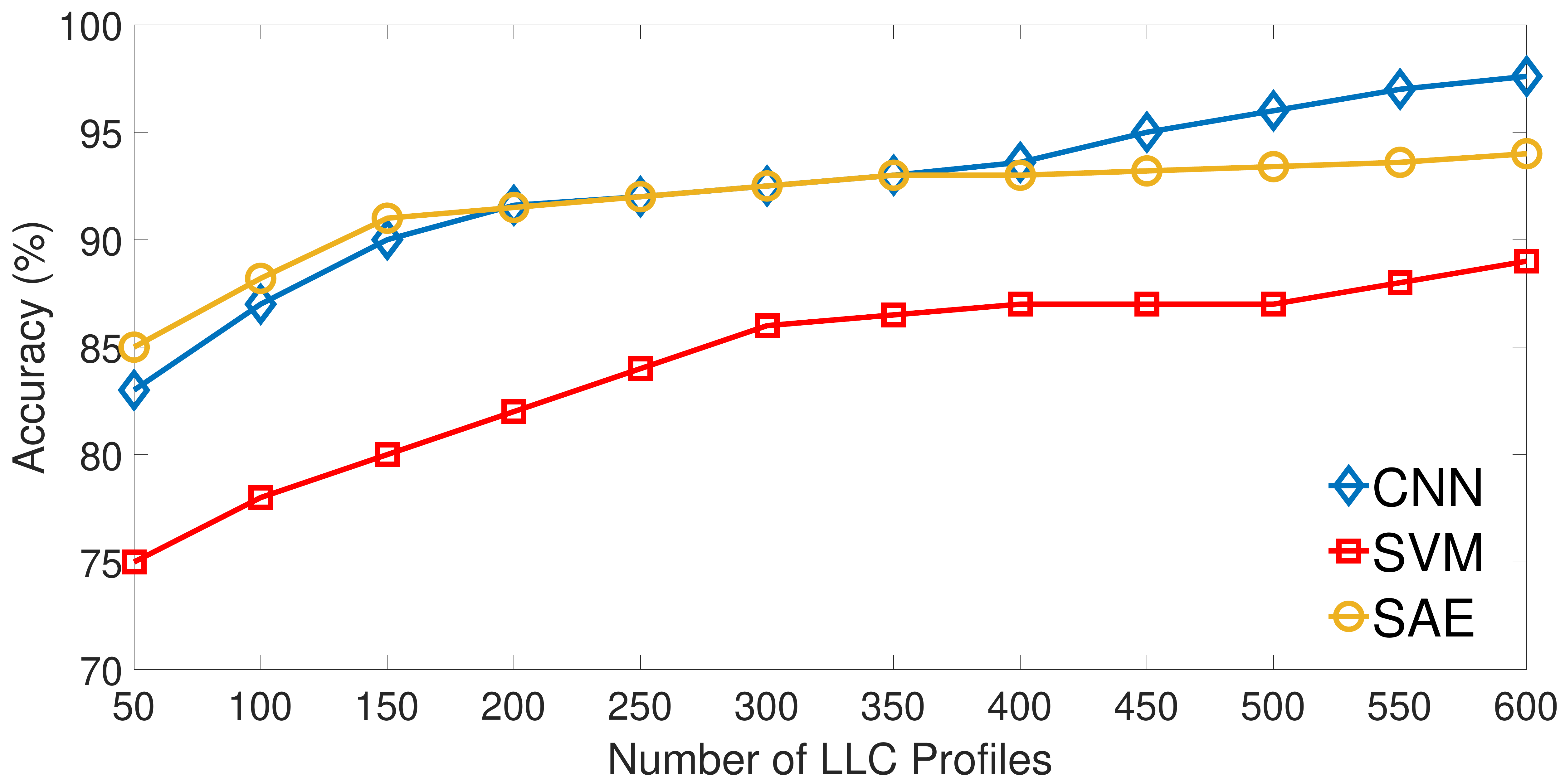}\label{fig:AFFT}}
  \caption{Classification results for application inference over an increasing number of LLC profiles for (a) Physical Cache Sets, (b) Virtual Cache Sets, and (c) Virtual Cache Sets + FFT feature vectors.}
  \label{fig:app_detection}
\end{figure}

A first comparison of ML/DL techniques and feature vectors is given in Figure~\ref{fig:app_detection}.
It contains three sub-plots that each show the classification results of SVM, SAE, and CNN over an increasing number of recorded LLC profiles.
90\% of the recorded profiles are used for training, whereas the rest is used to obtain the classification rates shown in the plots.
Note that in practice, recording one LLC profile on the target device is sufficient to conduct a successful inference attack.

Plot~\ref{fig:ACM} illustrates the results of the comparison attack, which is based on the \textit{physical cache set} feature vector.
The fine-grained LLC profiles allow all three classifiers to distinguish applications with high confidence.
CNN even achieves a classification rate of 97\%.
Plot~\ref{fig:ACMT} shows that classification rates drop, if the LLC profiles are based on virtual cache sets.
SAE even falls down to 80\%, while CNN remains above 90\%.
The CNN that we designed is therefore least affected by the simplified profiling technique of our inference attack.
The classification rates improve again, if the LLC profiles are transformed with an FFT, as shown in Plot~\ref{fig:AFFT}.
In particular, CNN and SAE benefit from this transformation, while SVM cannot fully leverage the information in the frequency spectrum.
Our CNN reaches a classification rate of 97.8\% and is thereby able to fully close the gap to the comparison attack.

\begin{figure}[t!]
  \centering
  \includegraphics[width=0.48\textwidth]{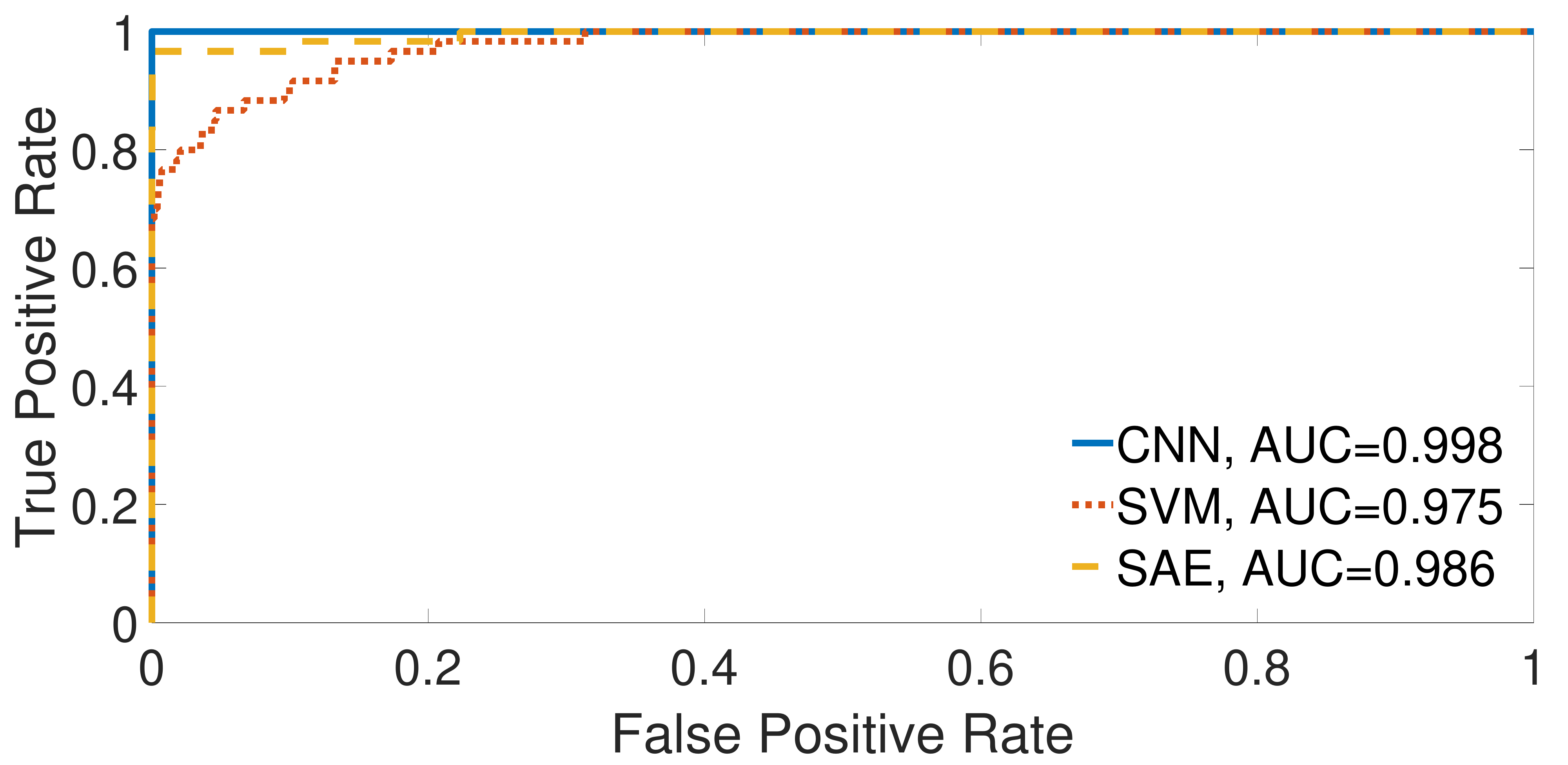}
  \caption{Average receiver operating characteristic (ROC) curves for SVM, SAE, CNN during application inference.}
  \label{fig:ROC_App}
\end{figure}

A further performance metric for the three ML/DL techniques is shown in Figure~\ref{fig:ROC_App}.
It displays the receiver operating characteristic (ROC) curves for the \textit{virtual cache set + FFT} feature vector.
The area under the curve (AUC) values are given in the plot legend.
The higher the AUC, the less the ML/DL technique suffers from false positives.
While all three classifiers produce low false positive rates, CNN outperforms SVM and SAE.
Based on the results of the application detection, we conclude that CNN is the most suitable classifier for our inference attack.
For website and video inference, we will therefore only present the CNN results.
The results of the other classifiers can be found in Appendix~\ref{sec:appendix}.

\begin{figure}[t!]
  \centering
  \includegraphics[width=0.48\textwidth]{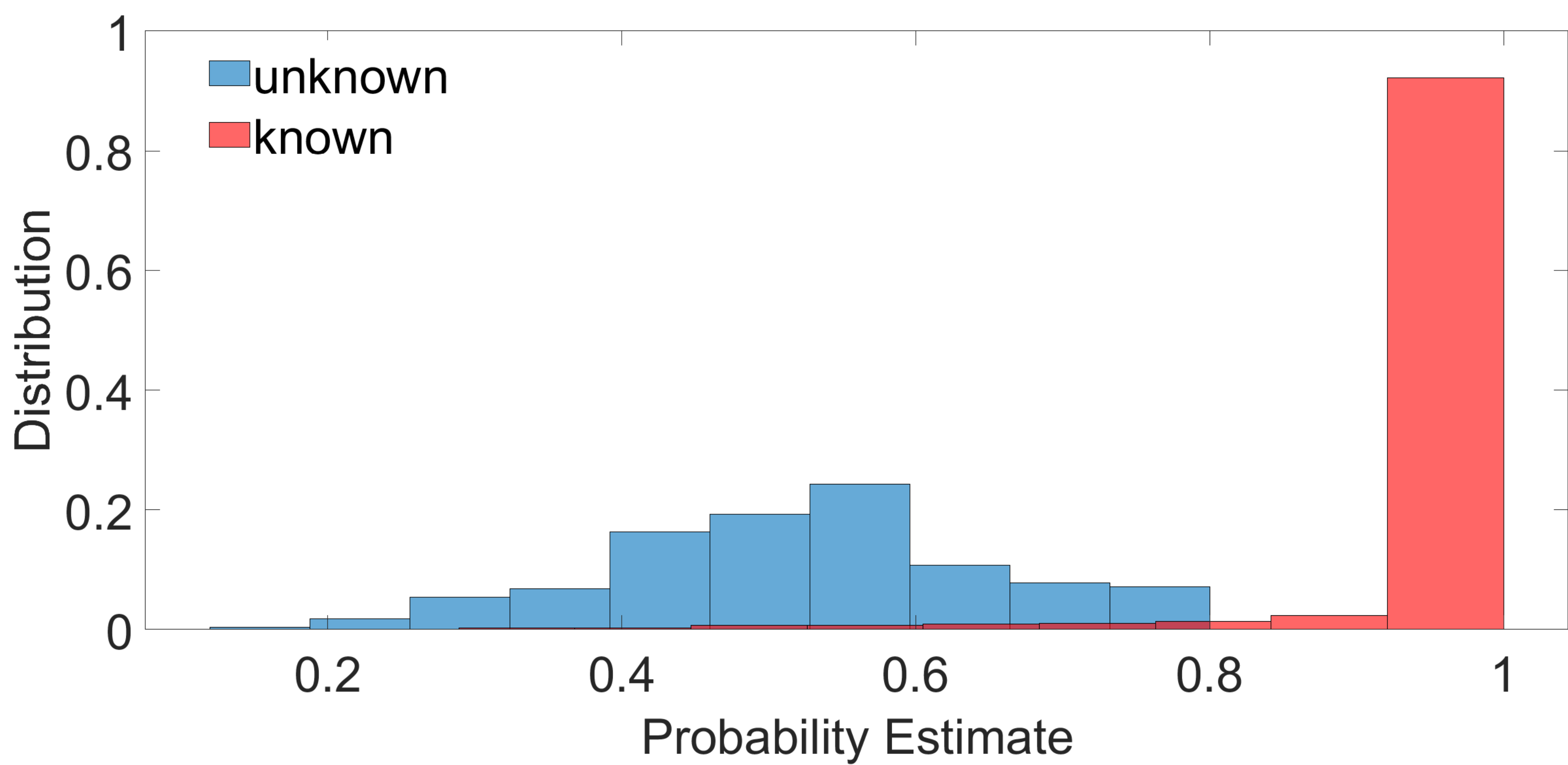}
  \caption{Probability estimates from the Softmax layer of our CNN while classifying known and unknown Apps.}
  \label{fig:unknown_App}
\end{figure}

\paragraph{\textbf{Unknown Applications.}}
The inherent nature of supervised learning is to recognize events that are similar to those used in the training phase.
In practice, however, events may occur that the model has never been trained with.
This also applies to our inference attack.
Naturally, we cannot train our models with all existing applications on the App store.
In fact, we want to focus only on Apps that are of interest.
Hence, we need a way to recognize and filter Apps we have not trained yet.
We achieve this by monitoring the probability estimates obtained from the Softmax layer.
Recall that we train only 70 Apps out of the 100 that are given in Table~\ref{tbl:appsdet} in Appendix~\ref{sec:appendix}.
When we classify all 100 Apps on our target device, we obtain the probability estimates shown in Figure~\ref{fig:unknown_App}.
All known Apps yield a high probability estimate close to 1, whereas unknown Apps yield estimates that are significantly lower.
We thus label each classification that yields a probability estimate below a threshold to be \textit{unknown}.
By setting this threshold to 0.84 in our experiments, we obtain false positive and false negative rates of 3\% and 1\%, respectively.
With this approach, our inference attack works reliably, even if unknown applications are used on the target device.

\subsubsection{Website Inference}

The results in the previous section illustrate that our malicious App can reliably detect running applications with high confidence.
Once a browser is detected, it is possible to infer the websites that are currently viewed.
We demonstrate this by profiling the LLC while visiting 100 different websites in Google Chrome.
The list of websites is given in Table~\ref{table:websites} in Appendix~\ref{sec:appendix}.
To emphasize that browsing histories are sensitive information, the list includes news, social media, political, and dating websites.
For each website, we profile the LLC for 1.5 seconds and again obtain $n_{T} = 96,000$ samples per virtual cache set.
The features vectors are constructed in the same way as for application detection.

\begin{figure}[t!]
  \centering
  \includegraphics[width=0.48\textwidth]{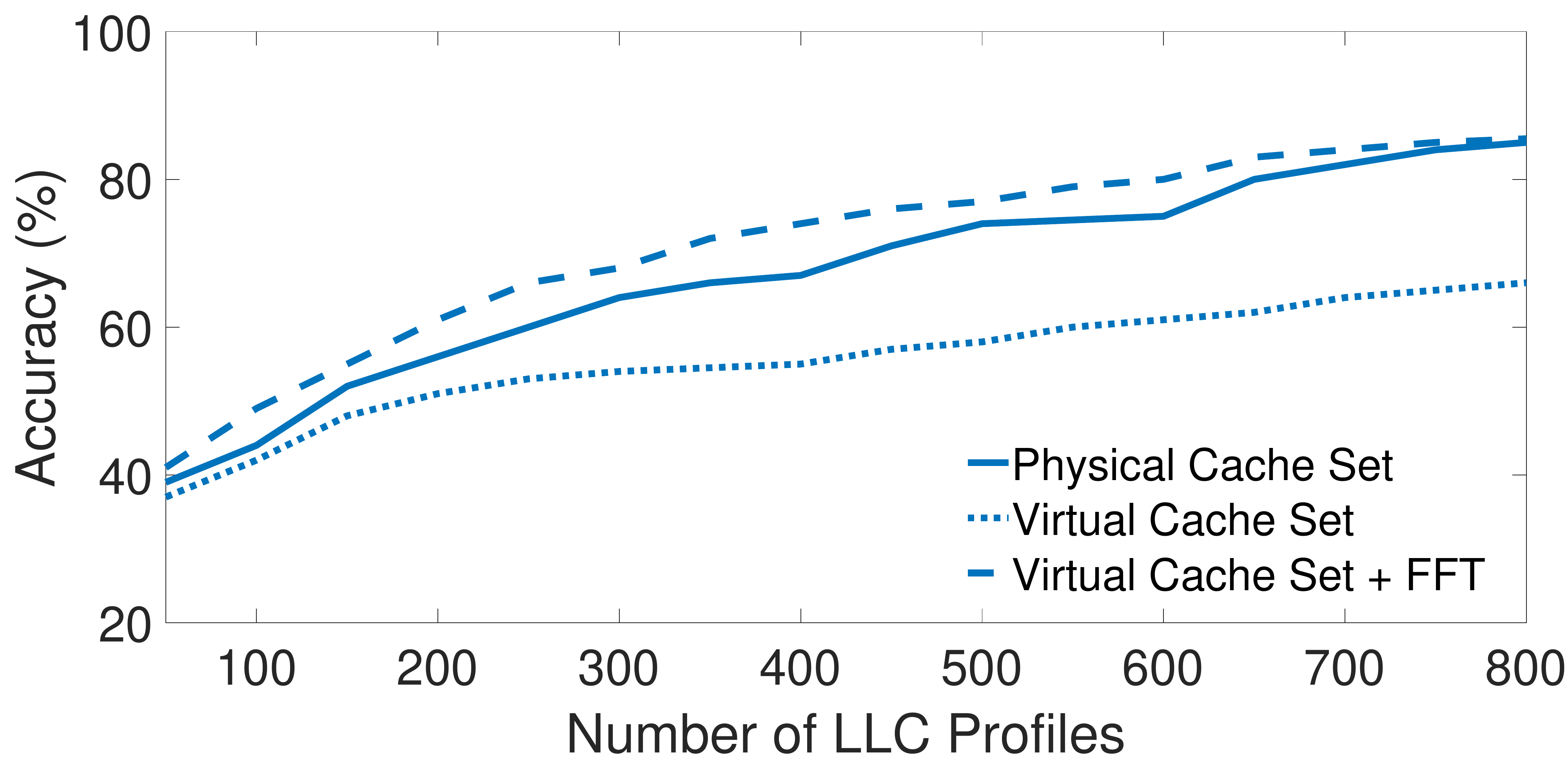}
  \caption{Classification results for website inference over an increasing number of LLC profiles for Physical (solid), Virtual (dotted), and Virtual Cache Set + FFT (dashed).}
  \label{fig:comparison_google_chrome}
\end{figure}

Figure~\ref{fig:comparison_google_chrome} shows the classification results for our CNN and all three feature vectors over an increasing number of LLC profiles.
Similar to application inference, the frequency transform of virtual cache set profiles matches and slightly overshoots the results of the comparison attack that is based on physical cache sets.
With a classification rate of 86\%, the CNN is able to infer viewed websites with satisfactory confidence.
The classification rate is lower compared to application inference, because loading and rendering websites leaves a weaker footprint in the LLC than opening Apps.

\begin{figure}[t!]
  \centering
  \includegraphics[width=0.48\textwidth]{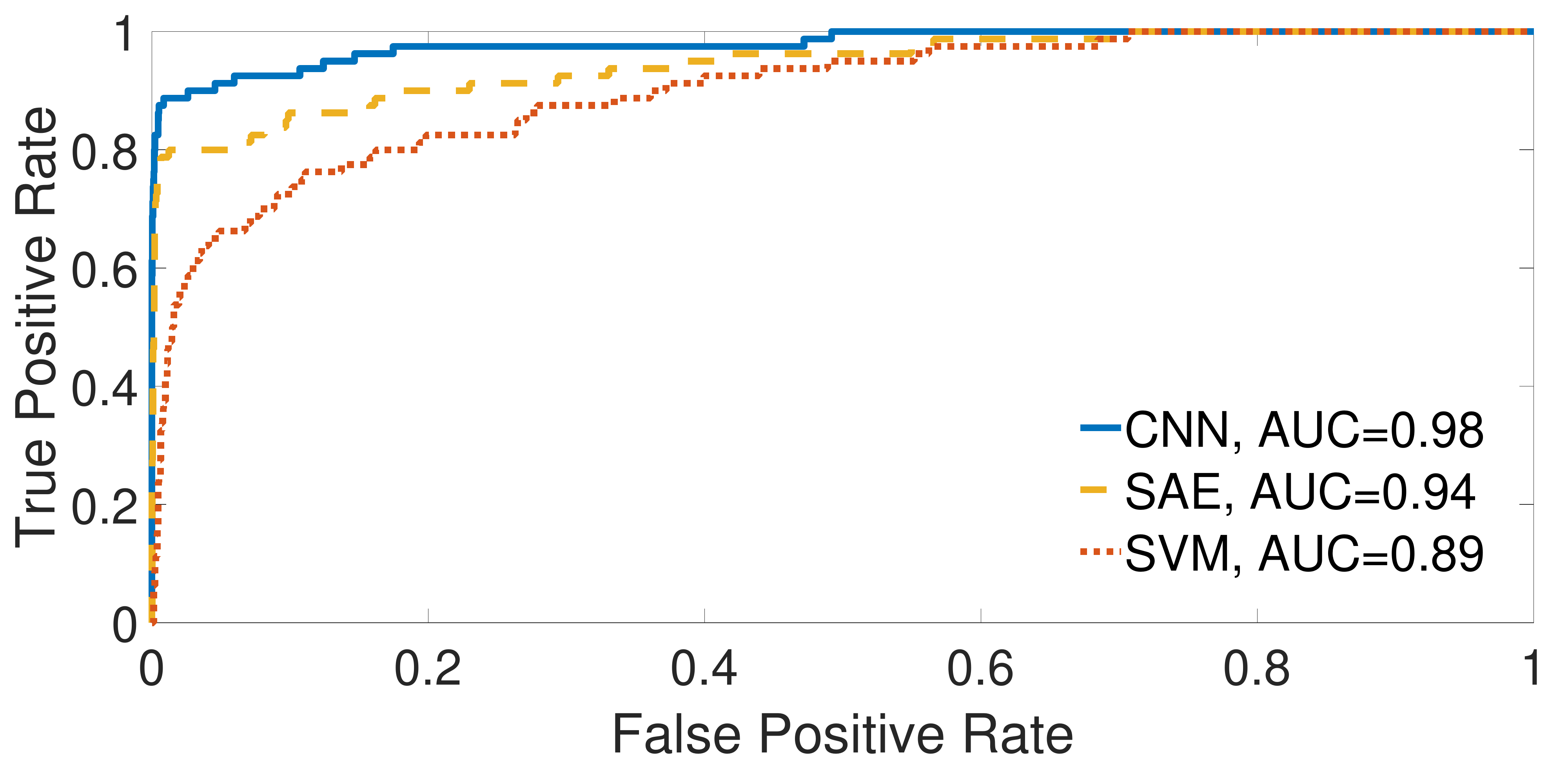}
  \caption{Average receiver operating characteristic (ROC) curves for SVM, SAE, CNN during website inference.}
  \label{fig:ROC_Website}
\end{figure}

The ROC curves for the virtual cache set + FFT feature vector are shown in Figure~\ref{fig:ROC_Website}.
The AUC values in the plot legend again illustrate that our CNN yields the lowest number of false positives.
This classifier and feature vector combination is the best choice for website inference.

\paragraph{\textbf{Unknown Websites.}}
As previously, we train the CNN with only 70 websites and subsequently classify all 100 websites from Table~\ref{table:websites}.
The probability estimates of the Softmax layer are similar to the application inference and therefore not shown here.
For a threshold of 0.74, we obtain false positive and false negative rates of 16\% and 3\%, respectively.
The inference attack is thus sufficiently robust against encountering unknown websites during profiling.

\subsubsection{Video Inference}

In addition to websites, we also infer videos that are being streamed in the Netflix and Youtube applications.
We profile a total of 20 videos, which are given in Table~\ref{table:videos} in Appendix~\ref{sec:appendix}.
In contrast to previous evaluations, we increase the profiling phase to 6 seconds.
This is because the LLC footprint of videos is less distinct compared to applications and websites.
Within the extended profiling phase, we collect $n_{T} = 384,000$ samples per virtual cache set.
For the FFT computation, the number of bins $n_{fft}$ is increased to 60.
Due to the high number of feature values, the size of the first Convolution layer in our CNN is increased to 1024.
The rest of the feature vectors are constructed in the same way as for application and website inference.

\begin{figure}[t!]
  \centering
  \includegraphics[width=0.48\textwidth]{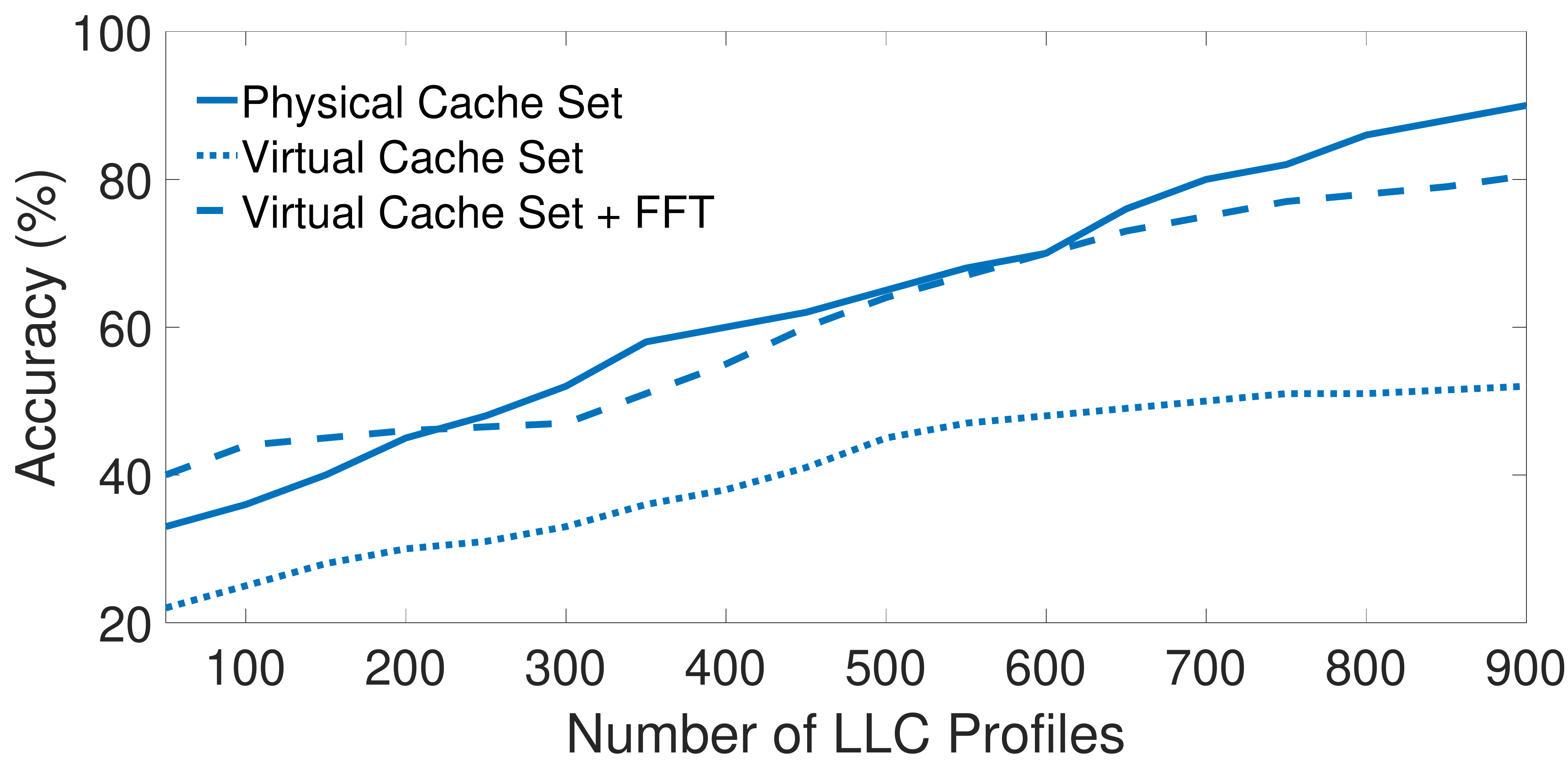}
  \caption{Classification results for video inference over an increasing number of LLC profiles for Physical (solid), Virtual (dotted), and Virtual Cache Set + FFT (dashed).}
  \label{fig:comparison_video}
\end{figure}

\begin{figure}[t!]
  \centering
  \includegraphics[width=0.48\textwidth]{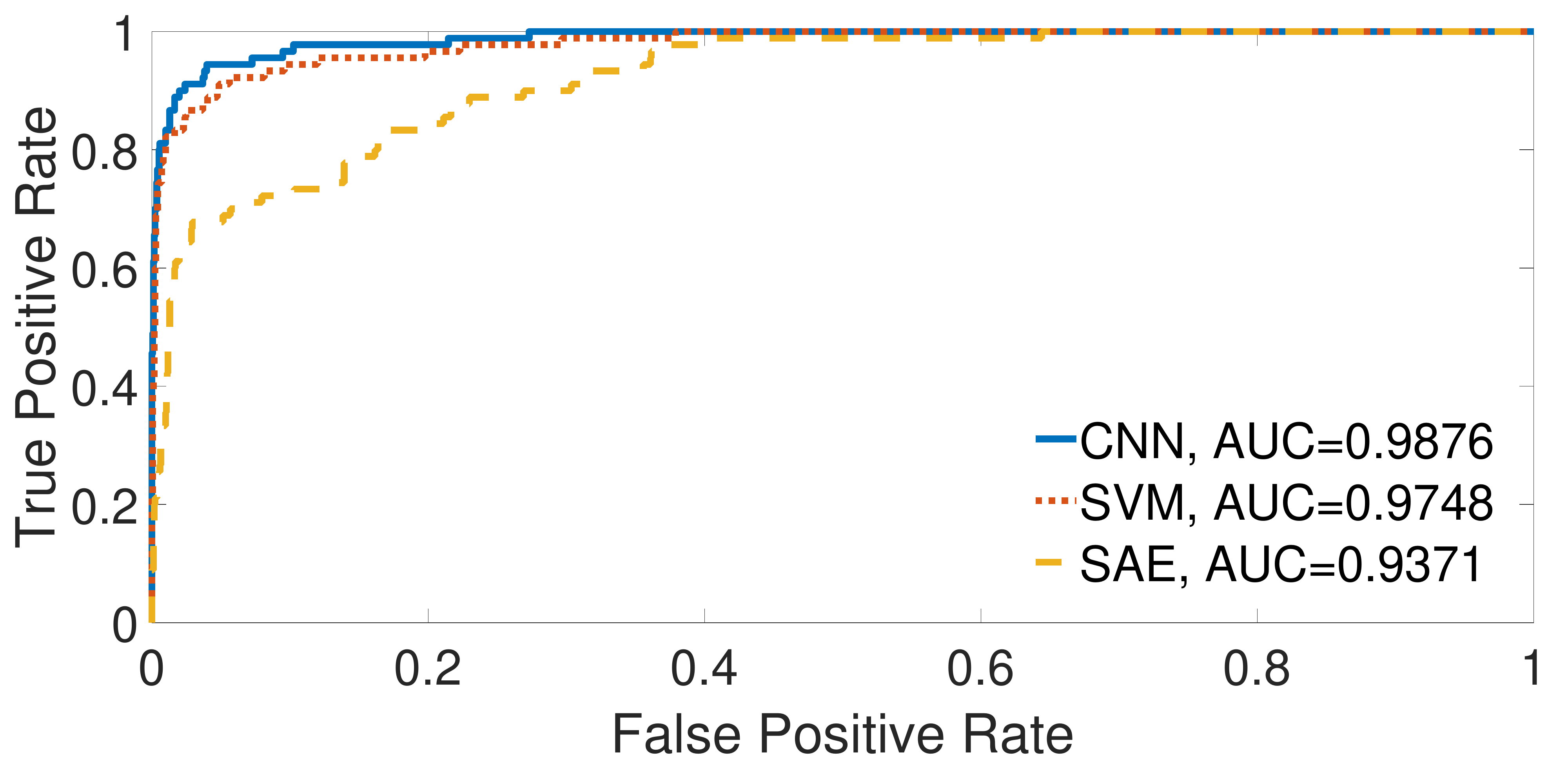}
  \caption{Average receiver operating characteristic (ROC) curves for SVM, SAE, CNN during video inference.}
  \label{fig:ROC_Video}
\end{figure}

Figure~\ref{fig:comparison_video} shows the classification results for our CNN and all three feature vectors over an increasing number of LLC profiles.
The FFT over the virtual cache sets again matches the comparison attack, but eventually falls behind by 10\%.
With a classification rate of 80\%, our inference attack is able to infer streamed videos with moderate success.
We believe that the LLC profiles do not contain enough information to distinguish multiple videos, as video processing is a rather homogeneous task.
In addition, parts of the video decoding are typically outsourced to the GPU, which further reduces the cache footprint.
Regarding the ROC curves, which are shown in Figure~\ref{fig:ROC_Video}, our CNN again outperforms SVM and SAE, with SVM not falling far behind.
Due to the reduced success rate for video classification, we skipped the evaluation of unknown videos.
Yet, we expect them to follow the same trend as for application and website inference.

\subsubsection{Discussion}

The previous sections show that modern machine learning techniques enable successful inference attacks even when simple cache profiling methods are employed.
Throughout our experiments, frequency-domain transforms of coarse-grained LLC profiles yield high success rates when classified with Convolutional Neural Networks.
This classification closely matches and sometimes outperforms the fine-grained comparison attack that is based on a more powerful attacker.
The FFT thereby reduces the noise in the measurements, while the shift invariance of CNNs compensates the lacking order in which LLC sets are profiled.
The limit of our inference attack becomes apparent when the LLC activity is less distinct or faints.
While applications and websites can reliably be inferred, the accuracy drops for video classification.
This could be improved by increasing the profiling time or including other side-channels such as GPU activity.
Nevertheless, the results clearly indicate that a carefully crafted and well-trained CNN enables inference attacks that are robust, easy to implement, and therefore practical.

\section{Related Work}
\label{sec:related_work}

The attack presented in this paper relates to previous work in the areas of website and application inference, cache attacks on ARM-based systems, and machine learning in the context of side-channel attacks (SCAs).
The following sub-sections discuss the relations to these areas in more detail.

\subsection{Website and Application Inference}

In literature, the inference of visited websites has been investigated from many perspectives.
Server-side website inference has been proposed through caching~\cite{FeltenSchneider2000} and rendering~\cite{LiangEtAl2014} of website elements, visited URL styles~\cite{JacksonEtAl2006}, user interactions~\cite{WeinbergEtAl2011}, quota management~\cite{KimEtAl2016}, and shared event loops~\cite{VilaKoepf2017}.
Panchenko et al.~\cite{panchenko2016website} use traffic analysis to detect visited websites in the Tor network.
Zhang et al.~\cite{zhang2018level} exploit iOS APIs to infer visited websites and running applications on mobile platforms.
Spreitzer et al.~\cite{spreitzer2018procharvester,spreitzer2018scandroid} obtain distinct features from the \textit{procfs} filesystem and use Android APIs to infer opened web pages and applications.
Lee et al.~\cite{lee2014stealing} exploit uninitialized GPU memory pages to detect websites, while Naghibijouybari et al.~\cite{naghibijouybari2018rendered} exploit OpenGL APIs and GPU performance counters for this task.
Gulmezoglu et al.~\cite{gulmezoglu2017perfweb} show that hardware performance events on modern processors give distinct features for websites visited in Google Chrome and Tor browsers.
Diao et al.~\cite{diao2016no} infer application information through system interrupts.
Jana and Shmatikov~\cite{JanaShmatikov2012} demonstrate that websites leave a distinct memory footprint in the browser application.
Hornby~\cite{Hornby2016} shows that this footprint can be observed from a malicious application via the processor cache.
Oren et al.~\cite{OrenEtAl2015} as well as Gruss et al.~\cite{gruss2015practical} demonstrate that opened websites and their individual elements can be inferred from cache observations taken from a malicious JavaScript applet.

Shusterman et al.~\cite{ShustermanEtAl2018} extend this work by inferring websites from JavaScript with simple last-level cache profiles that are classified by Convolutional Neural Networks and Long Short-Term Memory.
As this is concurrent work to ours\footnote{Published on Nov. 17$^{\text{th}}$: \url{https://arxiv.org/abs/1811.07153}.}, we provide a closer comparison later in the section.
Gulmezoglu et al.~\cite{gulmezoglu2017cache} use cache observations to detect running applications in co-located virtual machines.
In this work, we also use measurements of cache activity to infer running applications, visited websites, and streamed videos.
The comparison of our results with other attacks is given in Table~\ref{table:comparison}.
For website classification, GPU-, network traffic-, and operating system-based attacks achieve higher success rates than our inference attack.
In contrast to these attacks, we don't need access to GPU, network, or OS APIs, which can be restricted or easily monitored.
We rely on simple memory accesses and coarse-grained timing measurements, which are difficult to restrict and monitor.
Compared to the LLC-based attack in~\cite{OrenEtAl2015}, our success rates are higher, even though we classify significantly more websites.
Compared to the results by Shusterman et al.~\cite{ShustermanEtAl2018}, we achieve similar classification rates.
At the same time, we relax the attacker model by not only compensating imprecise timing sources but also random cache replacement policies.
For application detection, the success rate of our approach is, to the best of our knowledge, the highest one in literature.

\begin{table*}[t]
  \caption{Comparison of website and application inference attacks.}
  \centering
  \begin{tabular}{| c | c | c | c | c |}
    \hline
    Scenario & Attack & Attack Vector & Accuracy (\%) & \# of Classes \\ [0.5ex]
    \hline
    \multirow{11}{*}{Websites} & \textbf{This work} & \textbf{Last-level Cache (LLC)} & \textbf{85.8} & \textbf{70} \\
    & \cite{OrenEtAl2015} & LLC & 82.1 & 8 \\
    & \cite{ShustermanEtAl2018} & LLC & 86.1 & 100 \\
    & \cite{gulmezoglu2017perfweb} & CPU Performance Events & 84.0 & 30 \\
    & \cite{naghibijouybari2018rendered} & GPU Performance Events & 93.0 & 200 \\
    & \cite{lee2014stealing} &  Uninitialized GPU Memory Pages & 95.4 & 100 \\
    & \cite{JanaShmatikov2012} & Scheduling Statistics & 78.0 & 100 \\
    & \cite{VilaKoepf2017} & Shared Event Loops & 76.7 & 500 \\
    & \cite{panchenko2016website} & Traffic Analysis & 92.5 & 100 \\
    & \cite{zhang2018level} & iOS APIs & 68.5 & 100 \\
    & \cite{spreitzer2018scandroid} & Java-based Android API & 89.4 & 20 \\
    & \cite{spreitzer2018procharvester} & ProcFS Leaks & 94.0 & 20 \\\hline
    \multirow{6}{*}{Applications} & \textbf{This work} & \textbf{LLC} & \textbf{97.8} & \textbf{70} \\
    & \cite{gulmezoglu2017cache} & LLC & 78.5 & 40 \\
    & \cite{zhang2018level} & iOS APIs & 89.0 & 120 \\
    & \cite{spreitzer2018scandroid} & Java-based Android API & 85.6 & 20 \\
    & \cite{diao2016no} & Interrupt Handling Mechanism & 87.0 & 100 \\
    & \cite{spreitzer2018procharvester} & ProcFS Leaks & 96.0 & 100 \\\hline
  \end{tabular}
  \label{table:comparison}
\end{table*}

\paragraph{\textbf{Comparison with Shusterman et al.}}
In~\cite{ShustermanEtAl2018}, the authors present a similar inference attack than the one proposed in this work.
The LLC profiling is implemented with so-called one-dimensional \textit{memorygrams}.
Essentially, this is a buffer as large as the LLC, which is repeatedly accessed.
The time to access the entire buffer then relates to the activity in the LLC and is used to infer websites.
In contrast, we build eviction sets and group them to profile contiguous parts of the LLC.
This provides a more fine-grained view on the activity in the LLC.
Shusterman et al. profile the LLC for 30 seconds, while we achieve our results with a profiling phase of 1.5 seconds.
The authors further choose their CNN parameters based on the success rate, while we select the parameters based on validation loss.
Using the validation loss makes the training model more robust, because success rates increase with over-fitting.
The differences in parameter selection are essentially the number of Convolution layers, the kernel size, and the Pooling size.
We trained our model with 2 Convolution layers, which yield a lower validation loss compared to 3 layers.
The kernel size is kept the same in each Convolution layer, while in~\cite{ShustermanEtAl2018} it is varied with each layer.
Finally, we use a Pooling size of 2, while Shusterman et al. use 4.
One reason for these differences is that we incorporated the results by Prouff et al.~\cite{benadjilastudy} in addition to conducting preliminary experiments.
The classification rates for Google Chrome in~\cite{ShustermanEtAl2018} are comparable to the ones we achieve.
In summary, both Shusterman et al. and this work propose inference attacks that share a common goal, but differ regarding approach and attack environment.
In~\cite{ShustermanEtAl2018} the attack was launched from JavaScript on Intel CPUs, while we conducted our attacks on Android and ARM.
Yet, the achieved classification rates are similar.
We believe this emphasizes that inference attacks of this kind are a practical, cross-platform attack that need to be addressed, if we want to safeguard user privacy in the long-term.

\subsection{Cache Attacks on ARM}

In the past, the field of cache attacks has been predominantly focused on x86 processors.
Most attacks known today either use dedicated flush instructions~\cite{GullaschEtAl2011,YaromFalkner2014,GrussEtAl2016a} or targeted thrashing of cache sets~\cite{TromerEtAl2010,GrussEtAl2015,IrazoquiEtAl2015,LiuEtAl2015,GenkinEtAl2018} to manipulate and observe the processor cache.
These techniques have initially been demonstrated on ARM processors by Lipp et al.~\cite{LippEtAl2016}.
In contrast to Intel x86, ARM processors complicate attacks with random replacement policies, exclusive and non-inclusive cache hierarchies, and internal line locking mechanisms~\cite{GreenEtAl2017}.
In this work, we show that despite these challenges, simple observations of the last-level cache are sufficient to infer user activity on ARM-based mobile devices.
Unlike previous work, we pair these simple observations with advanced machine learning techniques and thereby overcome the attack difficulties on ARM processors.

\paragraph{Comparison with Lipp et al.}
In~\cite{LippEtAl2016}, the authors perform multiple cache attacks on ARM devices, including Prime+Probe~\cite{TromerEtAl2010}, the attack technique employed in this work.
For this reason, we conducted an evaluation that compares the Prime+Probe technique by Lipp et al. to ours.
In particular, we set up an experiment, in which we try to classify the first 20 websites from Table~\ref{table:websites} in Appendix~\ref{sec:appendix}.
We obtain the Prime+Probe code from the GitHub repository~\cite{GrussGithub} by Lipp et al., and run the eviction strategy evaluator on our test device.
The strategy 22-1-6 yields the highest eviction rate of 98\% on our ARM Cortex-A57.
The code by Lipp et al. uses \textit{pagemap} entries to find eviction sets (thus requiring root privileges), while we employ algorithms \ref{alg:evict} and \ref{alg:duplicate} that work without privileges.
The profiling phase for the website classification is 1.5 seconds.
We collect 800 LLC profiles for each website, and use 90\% as training data and the rest as test data.
For the comparison, we derive the \textit{physical cache set} and \textit{virtual cache set + FFT} feature vectors, as described in Section~\ref{sec:attack}.
We omit the \textit{virtual cache set} feature vector, as it yielded lower accuracies than the other ones throughout our experiments.
Table~\ref{tab:cratecomp} shows the obtained classification rates based on the CNN classifier.

\begin{table}[h!]
  \centering
  \caption{Classification rates comparison using CNN.}
  \label{tab:cratecomp}
  \begin{tabular}{c|c|c}
    \hline
    \textbf{Feature} & \textbf{Lipp et al.} & \textbf{This Work} \\ \hline
    Physical      & 90\% & 93\% \\
    Virtual + FFT & 85\% & 94\% \\
    \hline
  \end{tabular}
\end{table}

\noindent While classification rates are close when profiling individual cache sets, our approach achieves a higher success rate for virtual cache sets.
We believe this is because our eviction sets are smaller and therefore more resilient against background cache activity during the profiling.
In summary, our profiling technique achieves higher classification rates while requiring no root privileges to find eviction sets.

\subsection{Machine Learning and SCAs}

Side-channel attacks (SCAs) typically rely on signal processing and statistics to infer information from observations.
This closely relates them to machine learning techniques.
In 2002, template attacks~\cite{brumley2009cache, chari2002template} were first introduced, which use Naive Bayes classifiers to match leakage signals on training and target devices.
Since 2011, more advanced machine learning approaches were introduced to side-channel literature~\cite{hospodar2011machine}.
Lerman et al.~\cite{lerman2011side} use Random Forests (RFs), SVMs, and Self-Organizing Maps (SOMs) to analyze a 3DES implementation and to compare the effectiveness of ML techniques against template attacks.
Later, Heuser et al.~\cite{heuser2012intelligent} applied multi-class SVMs to classify multi-bit values, pointing out that SVM-based analysis is more robust against noisy environments than classical template attacks.
In the same year, Zhang et al.~\cite{ZhangEtAl2012} showed that multi-class SVMs can be used in combination with Hidden Markov Models (HMMs) to recover RSA decryption keys from cache traces in a cloud environment.
Gulmezoglu et al.~\cite{gulmezoglu2017cache} showed that SVM-based approaches can be used to extract features from FFT components obtained from LLC traces.
Even though SVM can perform better than template attacks, the main problem with SVM is the choice of regularization and kernel parameters, which is slow and costly.
In addition, when the number of features increases, finding meaningful boundaries gets more difficult in a non-linear scenario.
Neural Networks (NN) have also been introduced to side-channel attacks.
Martinasek et al.~\cite{martinasek2013innovative, martinasek2013optimization} showed that basic NN techniques can recover AES keys with a 96\% success rate.
With the increasing popularity of Deep Learning, corresponding techniques were also studied in SCAs.
In 2014, Zheng et al.~\cite{zheng2014time} used CNNs to classify time series data with high accuracy.
In 2015, Beltramelli~\cite{beltramelli2015deep} introduced Long-Short Term Memory (LSTM), a special DL technique, to classify 12 different keypads obtained from smart watch motion sensors.
In 2016, Maghrebi et al.~\cite{maghrebi2016breaking} compared four DL-based techniques (AE, CNN, LSTM and MLP) with template attacks while attacking an unprotected AES implementation using power consumption.
The results indicated that CNNs outperform template attacks thanks to their advanced feature extraction capability.
In 2017, Schuster et al.~\cite{schuster2017beauty} showed that encrypted streams can be used to classify videos with CNNs.
In the same year, Gulmezoglu et al.~\cite{gulmezoglu2017perfweb} used hardware performance events to classify websites visited on a personal computer using SVM and CNN techniques.

\paragraph{Comparison with Prouff et al.}
It is important to follow a systematic approach when choosing parameters of CNNs.
Prouff et al.~\cite{benadjilastudy} studied the parameter selection of MLP and CNN in the context of side-channel attacks.
There are in total 4 rules to follow according to their work.
The first rule states that Convolution layers have the same configuration in the same block.
The second rule is that Pooling layers have a dimension of 2.
The third rule is that the number of filters in a Convolution layer is higher than those of the previous layer.
The fourth rule states that all Convolution layers have the same kernel size.
While we implement rules 1, 2 and 4, rule number 3 does not apply to our experiments.
Instead, the number of filters decreases for each Convolution layer.
In addition, we do not exhaustively explore batch sizes and optimization methods, since they do not significantly affect validation loss in our case.
During kernel size selection, we found that higher kernel sizes give better results, which is the opposite of the findings in~\cite{benadjilastudy}.
Since we are not conducting classical side-channel attacks, we evaluate our parameters based on the validation loss, while Prouff et al. find the best parameters based on the mean rank of the key classes.

\vspace*{-.3em}
\section{Conclusion}
\label{sec:conclusion}
\vspace*{-.2em}

Inference attacks undermine our privacy by revealing our most secret interests, preferences, and attitudes.
Unfortunately, modern processors, which constitute the core of our digital infrastructure, are particularly vulnerable to these attacks.
Footprints in the processor cache allow the inference of running applications, visited websites, and streamed videos.
Above all, the advances in machine learning, especially the concepts behind deep learning, significantly lower the bar of successfully implementing inference attacks.
Our work demonstrates that it is possible to execute an inference attack without privileges, permissions, or access to special programming interfaces and peripherals.
The simple nature of the attack code renders a detection on the target device next to impossible.
This simplicity is paired with the careful application of deep learning.
Interferences such as measurement noise, misalignment, or unknown and unfavorable processor features are thereby conveniently compensated.
The comparison with concurrent work furthermore indicates that inference attacks of this kind are ubiquitous and succeed across runtime environments and processing hardware.
For applications that value the privacy of their users, protection against inference attacks must thus be deployed in a timely manner.
A comprehensive solution, however, seems to require a closer collaboration between hardware manufacturers, operating system designers, and application developers.

%
%
\vspace*{-.3em}
\section{Acknowledgments}
\vspace*{-.2em}
This work is supported by the National Science Foundation, under grants CNS-1618837 and CNS-1814406.
Berk Gulmezoglu is also supported by the WPI PhD Global Research Experience Award 2017.

%
%
{\normalsize \bibliographystyle{acm}
\bibliography{refs}}

%
%
\newpage
\appendix
\section{Appendix}
\label{sec:appendix}

This section provides additional results regarding our experiments.
It lists the profiled applications, websites, and videos, and gives the parameters that were explored while constructing our Convolutional Neural Network.

\subsection{Website Inference}

\begin{figure}[t!]
  \centering
  \subfigure[Physical Cache Sets.]{\includegraphics[width=0.48\textwidth,clip,trim=0 2cm 0 0]{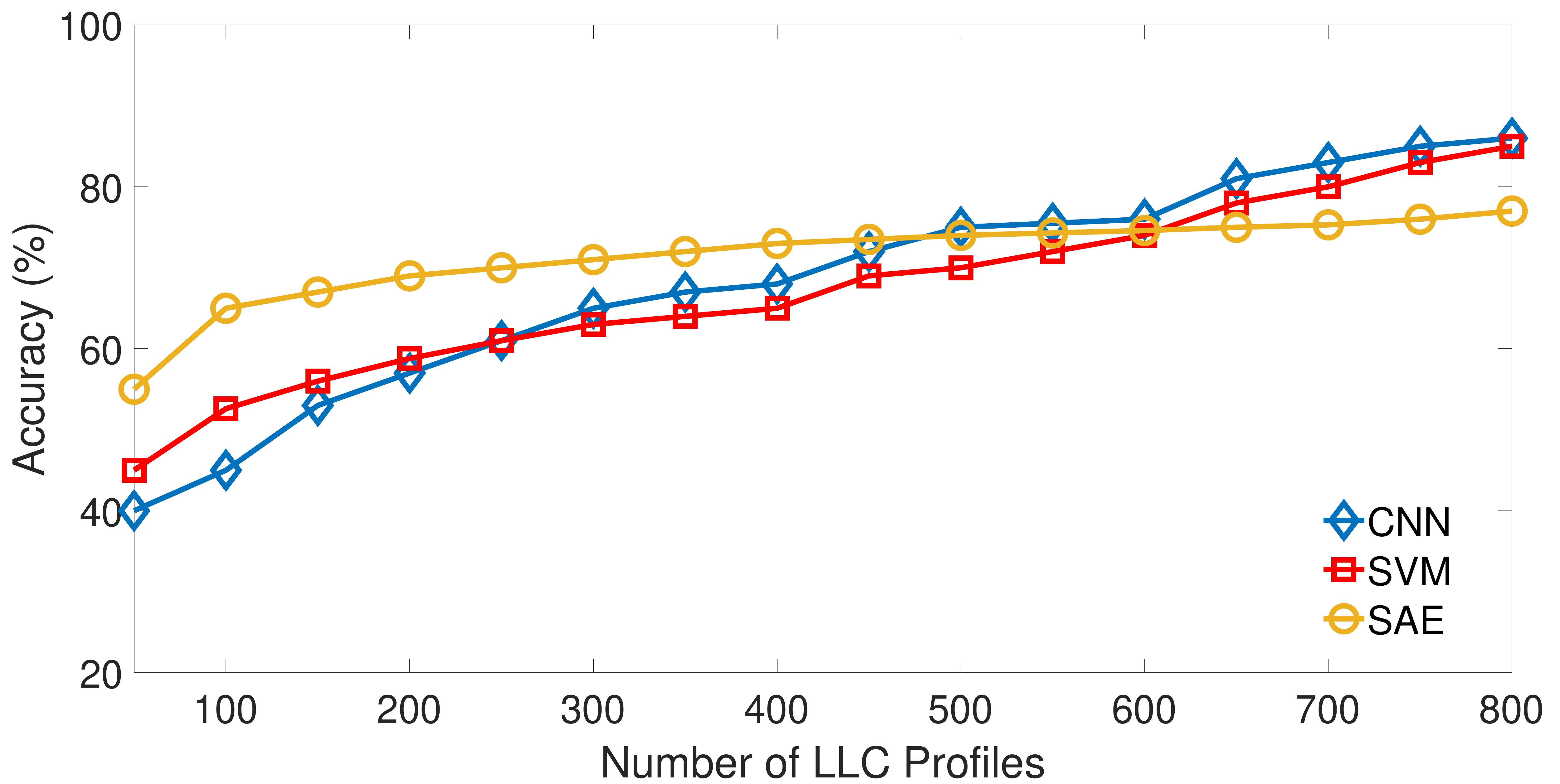}}
  \subfigure[Virtual Cache Sets.]{\includegraphics[width=0.48\textwidth,clip,trim=0 2cm 0 0]{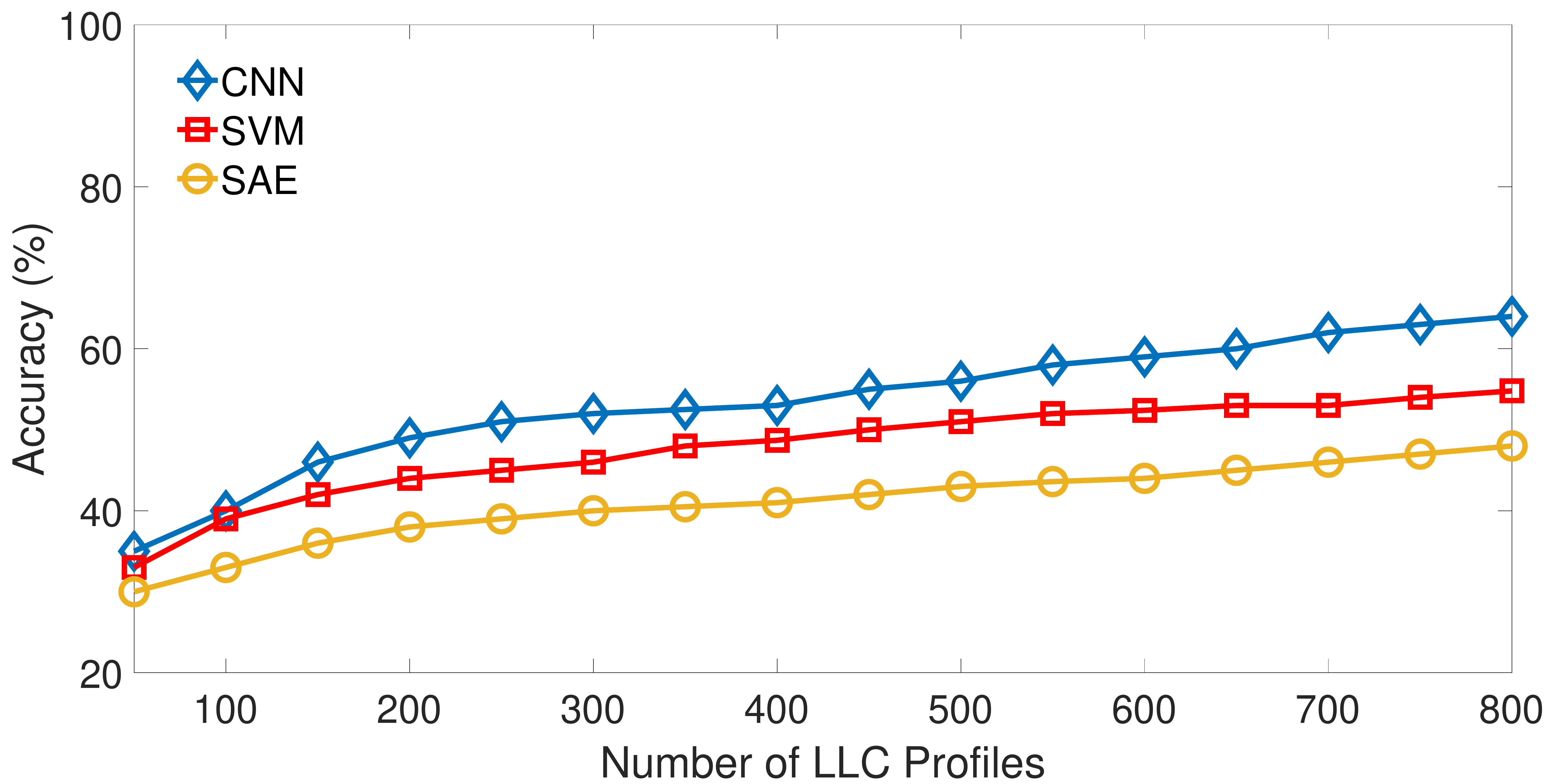}}
  \subfigure[Virtual Cache Sets + FFT.]{\includegraphics[width=0.48\textwidth]{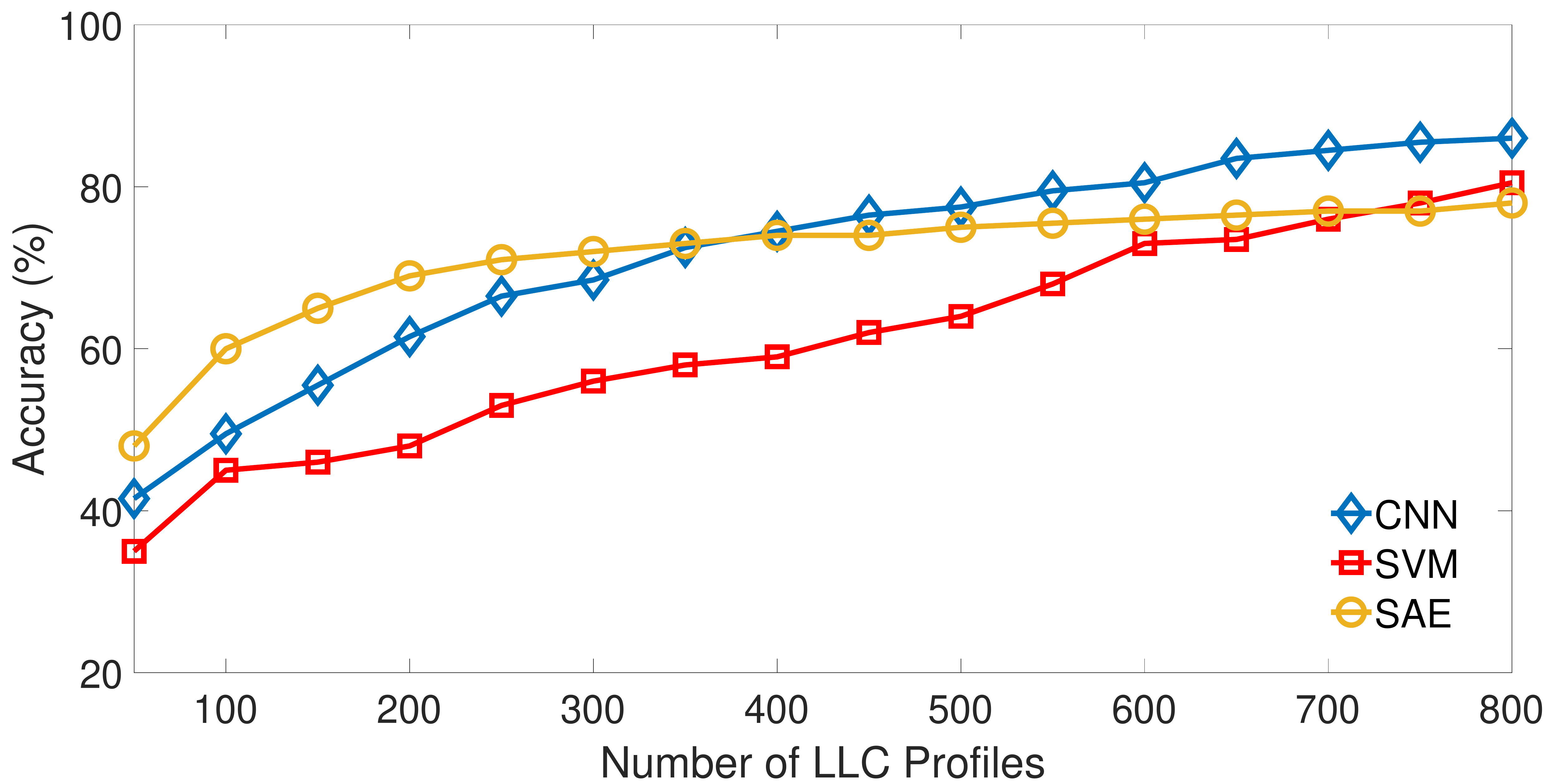}}
  \caption{Classification results for website inference over an increasing number of LLC profiles for (a) Physical Cache Sets, (b) Virtual Cache Sets, and (c) Virtual Cache Sets + FFT feature vectors.}
  \label{fig:GC}
\end{figure}

Figure~\ref{fig:GC} contains three sub-plots that each show the classification results of SVM, SAE, and CNN over an increasing number of recorded LLC profiles.
Similar to application inference, the classifiers are able to leverage the fine-grained information contained in the physical cache set feature vector.
While success rates drop for the virtual cache set feature vector, they recover for the virtual cache set + FFT feature vector.
In all three cases, the Convolutional Neural Network achieves the best classification rates.

\subsection{Video Inference}

\begin{figure}[t!]
  \centering
  \subfigure[Physical Cache Sets.]{\includegraphics[width=0.48\textwidth,clip,trim=0 2cm 0 0]{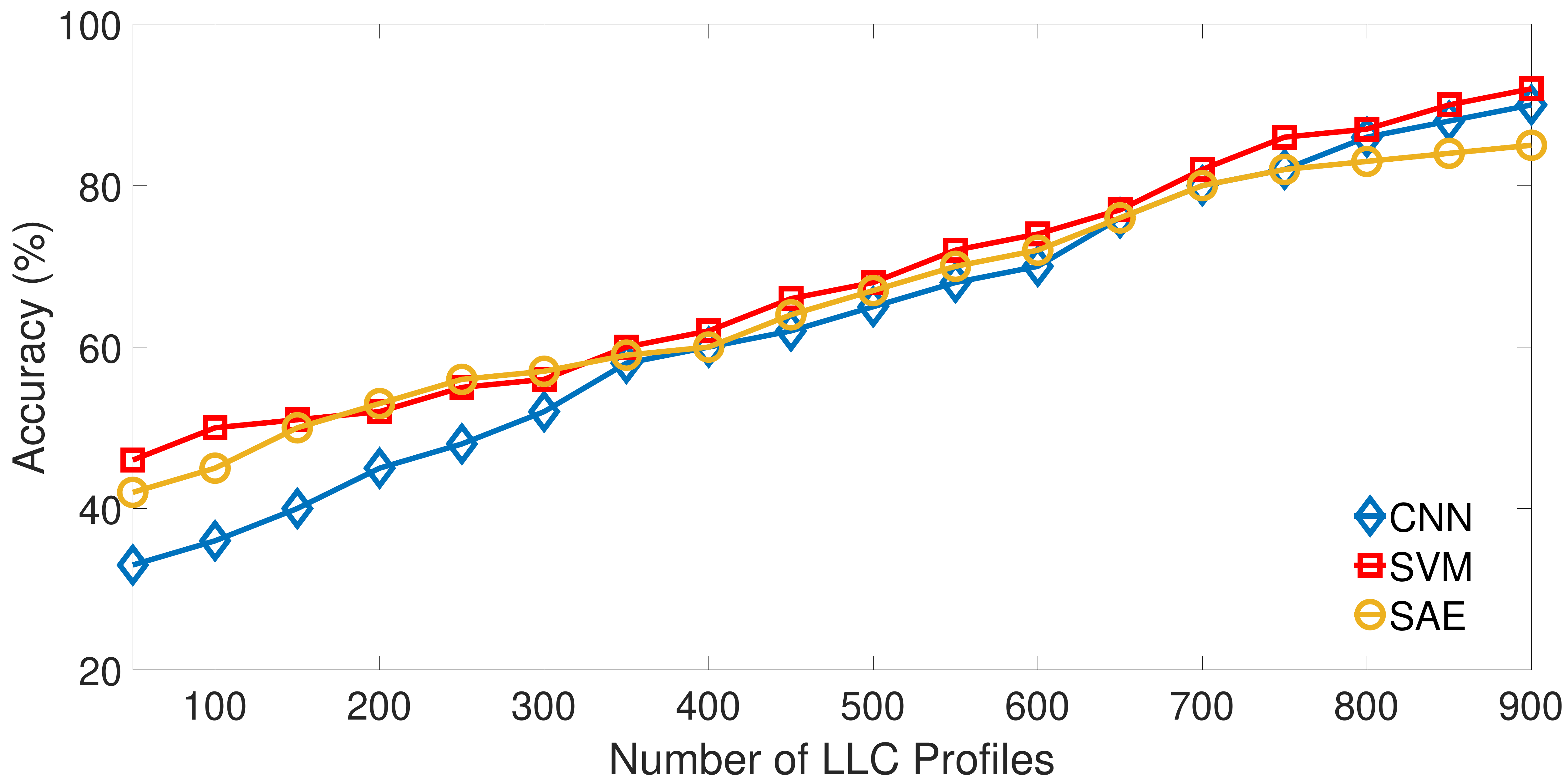}}
  \subfigure[Virtual Cache Sets.]{\includegraphics[width=0.48\textwidth,clip,trim=0 1.6cm 0 0]{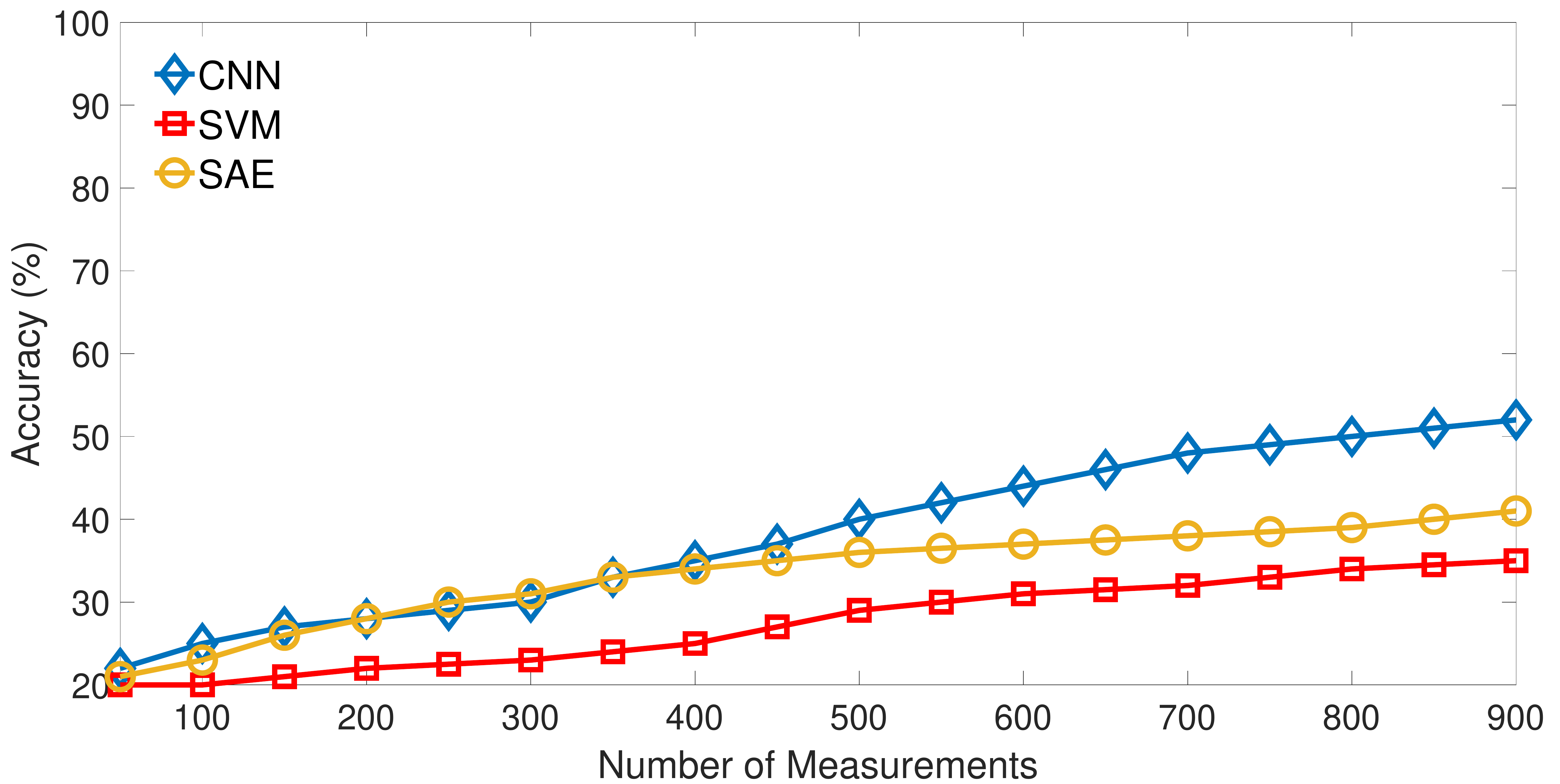}}
  \subfigure[Virtual Cache Sets + FFT.]{\includegraphics[width=0.48\textwidth]{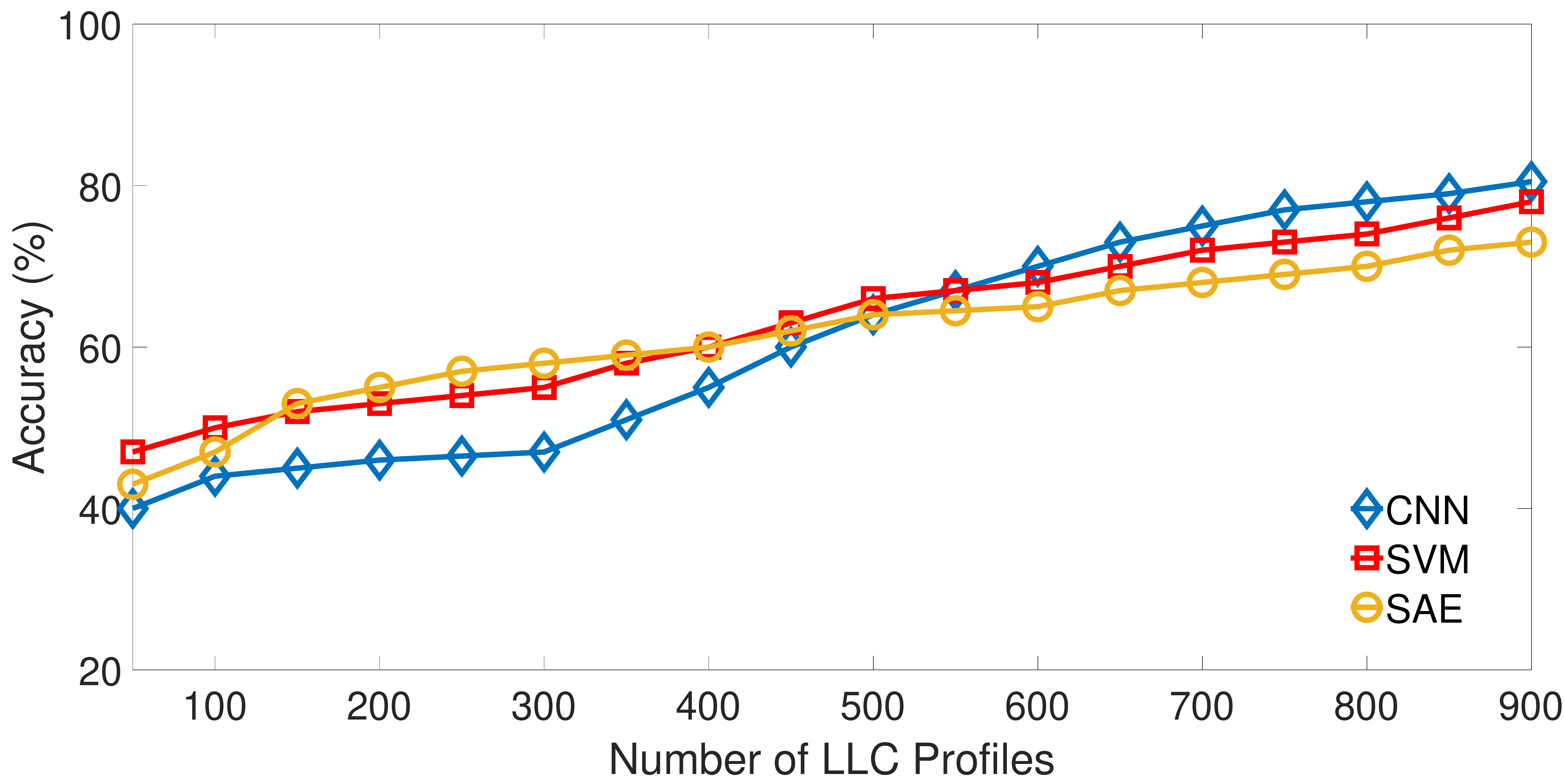}}
  \caption{Classification results for video inference over an increasing number of LLC profiles for (a) Physical Cache Sets, (b) Virtual Cache Sets, and (c) Virtual Cache Sets + FFT feature vectors.}
  \label{fig:GCNET}
\end{figure}

A similar trend can be observed in Figure~\ref{fig:GCNET}.
While the physical cache set feature vector allows confident classification, success drops significantly for virtual cache sets.
In contrast to application and website inference, the classification rates do not fully recover for the virtual cache set + FFT feature vector.
As discussed in Section~\ref{sec:results}, the LLC footprints while viewing different videos are not distinct enough to confidently infer them in this case.

\subsection{Profiled Applications, Websites, Videos}

Table~\ref{table:videos} lists the videos that are profiled in our experiments.
Youtube videos are chosen from the list of most watched videos on Youtube.
Trailers and recaps viewed in the Netflix App are from the series \textit{The House of Cards}.

\begin{table}[h!]
  \centering
  \caption{List of profiled videos.}
  \label{table:videos}
  \begin{tabular}{@{}ll@{}}
    \toprule
    \multicolumn{2}{c}{\textbf{Youtube (left) and Netflix (right) Videos}}\\ \midrule
    1) Despacito     & 1) Season 1 Trailer\\
    2) See You Again & 2) Season 2 Trailer\\
    3) Shape of You  & 3) Season 3 Trailer\\
    4) Gangnam Style & 4) Season 4 Trailer\\
    5) Uptown Funk   & 5) Season 5 Trailer\\
    6) Sorry         & 6) Season 1 Recap\\
    7) Sugar         & 7) Season 2 Recap\\ 
    8) Shake it Off  & 8) Season 3 Recap\\
    9) Roar          & 9) Season 4 Recap\\
    10) Bailando     & 10) Season 1 Trailer (Extended)\\
  \end{tabular}
\end{table}

\noindent Tables~\ref{tbl:appsdet} and~\ref{table:websites} list the profiled applications and websites.
Websites are thereby taken from the Alexa ranking~\cite{AlexaInternet2018}.

\clearpage

\clearpage

\begin{table}[t!]
  \centering
  \caption{List of profiled applications.}
  \label{tbl:appsdet}
  \begin{tabular}{@{}lll@{}}
    \toprule
    \multicolumn{3}{c}{\textbf{Applications}}\\ \midrule
    1) Spotify         &   35) Reddit             & 69) OurTime\\
    2) Snapchat        &   36) Imdb               & 70) HowAboutWe\\
    3) Instagram       &   37) Creditkarma        & 71) Tiktok \\    
    4) Facebook        &   38) Alexa              & 72) Canva \\
    5) Youtube         &   39) Yahoo              & 73) Autolist \\
    6) Chrome          &   40) Starz              & 74) Sephora\\
    7) Netflix         &   41) Zedge              & 75) Indeed\\
    8) Uber            &   42) Textnow            & 76) Marvel\\
    9) Twitter         &   43) Soundcloud         & 77) Hinge\\
    10) Bitmoji        &   44) Booking            & 78) Daylio\\
    11) Google Drive   &   45) Duolingo           & 79) Roku\\
    12) Pandora        &   46) Tinder             & 80) Investing\\
    13) NY Times       &   47) Joom               & 81) Ifood\\
    14) Pinterest      &   48) Xbox               & 82) Fitbit\\
    15) Lyft           &   49) Shazam             & 83) Goodrx\\
    16) InBrowser      &   50) Chase              & 84) Fastnews\\
    17) Firefox Focus  &   51) Huffington         & 85) Touchnote\\
    18) Orfox          &   52) Breitbart          & 86) Nike\\
    19) Musical Focus  &   53) Earspy             & 87) Sony\\
    20) Wish           &   54) Ispy               & 88) Kayak\\
    21) Hulu           &   55) Spycamera          & 89) Expedia\\
    22) Workout        &   56) Mspy               & 90) Sketch\\
    23) Waze           &   57) Secretagent        & 91) PlutoTV\\
    24) Walmart        &   58) Politico           & 92) Grubhub\\
    25) Wholefoods     &   59) TheHill            & 93) McDonald's\\
    26) Dairy Queen    &   60) Dailykos           & 94) Target\\
    27) Discord        &   61) Infowars           & 95) Trivia\\
    28) Venmo          &   62) Match              & 96) Starbucks\\
    29) Groupon        &   63) PlentyofFish       & 97) Horoscope\\
    30) Twitch         &   64) Zoosk              & 98) Beetles\\
    31) Yelp           &   65) eHarmony           & 99) Glasdoor\\
    32) Letgo          &   66) Okcupid            & 100)Tickermaster\\
    33) Iheart         &   67) Badoo              \\      
    34) eBay           &   68) Christian Mingle   \\
  \end{tabular}
\end{table}

\begin{table}[t!]
  \centering
  \caption{List of profiled websites.}
  \label{table:websites}
  \begin{tabular}{@{}lll@{}}
    \toprule
    \multicolumn{3}{c}{\textbf{Websites}} \\ \midrule
    1) Google         & 35) Hulu             & 69) BlackPeopleMeet\\
    2) Facebook       & 36) Quora            & 70) HowAboutWe\\
    3) Wikipedia      & 37) Salesforce       & 71) Oracle \\     
    4) Amazon         & 38) Wells            & 72) Reuters \\
    5) Reddit         & 39) Bank of America  & 73) BBC \\
    6) Yahoo          & 40) Stackoverflow    & 74) Nasa\\
    7) Twitter        & 41) Guardian         & 75) Eventbrite\\
    8) eBay           & 42) Forbes           & 76) Dailymotion\\
    9) Netflix        & 43) Dropbox          & 77) Blogger\\
    10) Linkedin      & 44) Mozilla          & 78) Nature\\
    11) Office        & 45) Soundcloud       & 79) Digg\\
    12) Cnn           & 46) Weebly           & 80) Wiley\\
    13) Espn          & 47) Vimeo            & 81) Wired\\
    14) Wikia         & 48) Adobe            & 82) Ted\\
    15) Twitch        & 49) Wordpress        & 83) Feedburner\\
    16) Live          & 50) Tumblr           & 84) Oath\\
    17) Instagram     & 51) Huffington       & 85) Ietf\\
    18) Craigslist    & 52) Breitbart        & 86) Nginx\\
    19) Paypal        & 53) Drudgereport     & 87) Springer\\
    20) Apple         & 54) Politico         & 88) Apache\\
    21) Bing          & 55) The Hill         & 89) Flickr\\
    22) Chase         & 56) Slate            & 90) Grawatar\\
    23) Zillow        & 57) Dailykos         & 91) Sourceforge\\
    24) Walmart       & 58) Infowars         & 92) Archive\\
    25) Yelp          & 59) Salon            & 93) Go\\
    26) Github        & 60) TheBlaze         & 94) Wix\\
    27) NY Times      & 61) Match            & 95) Myspace\\
    28) Pinterest     & 62) Plenty of Fish   & 96) Mysql\\
    29) Imdb          & 63) Zoosk            & 97) Time\\
    30) Microsoft     & 64) Okcupid          & 98) Cnbc\\
    31) Msn           & 65) eHarmony         & 99) Skype\\
    32) Fox News      & 66) Badoo            & 100) Alibaba\\
    33) Blogspot      & 67) Christian Mingle \\
    34) Dailymail     & 68) OurTime	        \\
  \end{tabular}
\end{table}

\clearpage

\begin{table*}[t]
  \caption{CNN parameter exploration. Final selection highlighted in bold.}
  \centering
  \label{table:parameter_selection}
  \begin{tabular}{| c | c | c | c | c | c | c | c | c | }
    \hline
    \multirow{2}{*}{\textbf{Convolution}} & \textbf{Max} & \multirow{2}{*}{\textbf{Dropout}} & \textbf{Kernel} & \multirow{2}{*}{\textbf{Dense}} & \textbf{Training} & \textbf{Training} & \textbf{Validation} & \textbf{Validation} \\
    & \textbf{Pooling} & & \textbf{Size} & & \textbf{Loss} & \textbf{Accuracy (\%)} & \textbf{Loss} & \textbf{Accuracy (\%)} \\
    \hline
    \multirow{30}{*}{} 1024 & 2 & 0.1 & 3 & 50 & 0.2568 & 93.54 & 0.7092 & 83.04 \\\hline
    512 & 2 & 0.1 & 3 & 50 & 0.2085 & 94.23 & 0.6876 & 84.03\\\hline
    256 & 2 & 0.1 & 3 & 50 & 0.2554 & 95.01 & 0.7127 & 82.75\\\hline
    128 & 2 & 0.1 & 3 & 50 & 0.2666 & 93.56 & 0.7307 & 82.78\\\hline
    64  & 2 & 0.1 & 3 & 50 & 0.2790 & 92.31 & 0.7342 & 82.68\\\hline
    32  & 2 & 0.1 & 3 & 50 & 0.5443 & 83.23 & 0.8038 & 80.56\\\hline
    16  & 2 & 0.1 & 3 & 50 & 0.3821 & 89.04 & 0.6910 & 82.38\\\hline
    8   & 2 & 0.1 & 3 & 50 & 0.4513 & 86.94 & 0.7057 & 82.29\\\hline
    512-256 & 2 & 0.1 & 3 & 50 & 0.2581 & 92.17 & 0.6436 & 84.40\\\hline
    512-128 & 2 & 0.1 & 3 & 50 & 0.3725 & 89.09 & 0.6510 & 84.18\\\hline
    512-64  & 2 & 0.1 & 3 & 50 & 0.6743 & 81.12 & 0.7638 & 80.93\\\hline
    512-32  & 2 & 0.1 & 3 & 50 & 0.4495 & 87.25 & 0.7355 & 81.46\\\hline
    512-256-128 & 2 & 0.1 & 3 & 50 & 0.2564 & 91.24 & 0.6440 & 84.55\\\hline
    512-256-64  & 2 & 0.1 & 3 & 50 & 0.3345 & 90.15 & 0.6609 & 84.09\\\hline
    512-256-32  & 2 & 0.1 & 3 & 50 & 0.3259 & 90.75 & 0.6984 & 81.25\\\hline
    512-256 & 2 & 0.1 & 3 & 50 & 0.2581 & 92.17 & 0.6436 & 84.40\\\hline
    512-256 & 4 & 0.1 & 3 & 50 & 0.4823 & 86.48 & 0.7467 & 82.45\\\hline
    512-256 & 8 & 0.1 & 3 & 50 & 0.5642 & 84.24 & 0.7160 & 81.54\\\hline
    512-256 & 2 & 0.2 & 3 & 50 & 0.2581 & 92.17 & 0.6436 & 84.80\\\hline
    512-256 & 2 & 0.3 & 3 & 50 & 0.2756 & 91.24 & 0.6783 & 83.34\\\hline
    512-256 & 2 & 0.4 & 3 & 50 & 0.2894 & 90.57 & 0.6928 & 82.86\\\hline
    512-256 & 2 & 0.5 & 3 & 50 & 0.3184 & 88.37 & 0.7293 & 81.43\\\hline
    512-256 & 2 & 0.2 & 6 & 50 & 0.4068 & 87.65 & 0.6583 & 83.45\\\hline
    512-256 & 2 & 0.2 & 9 & 50 & 0.3686 & 89.48 & 0.6314 & 85.21\\\hline
    512-256 & 2 & 0.2 & 18& 50 & 0.3079 & 91.00 & 0.6794 & 84.82\\\hline
    512-256 & 2 & 0.2 & 27& 50 & 0.3283 & 90.06 & 0.6915 & 83.87\\\hline
    512-256 & 2 & 0.2 & 9 & 100& 0.3387 & 90.03 & 0.6836 & 83.07\\\hline
    512-256 & 2 & 0.2 & 9 & 150& 0.3208 & 90.39 & 0.6456 & 83.57\\\hline
    \rowcolor{gray!30}
    \textbf{512-256} & \textbf{2} & \textbf{0.2} & \textbf{9} & \textbf{200}& \textbf{0.3104} & \textbf{91.25} & \textbf{0.6218} & \textbf{85.76}\\\hline
    512-256 & 2 & 0.2 & 9 & 250& 0.3487 & 89.74 & 0.6424 & 83.38\\\hline
    512-256 & 2 & 0.2 & 9 & 300& 0.3562 & 88.96 & 0.6592 & 82.51\\\hline
    512-256 & 2 & 0.2 & 9 & 350& 0.3859 & 86.52 & 0.6834 & 82.15\\\hline
  \end{tabular}
\end{table*}

\end{document}